\documentclass[fleqn,usenatbib]{mnras}

\usepackage[T1]{fontenc}
\usepackage{ae,aecompl}
\usepackage{graphicx}	
\usepackage{amsmath}	
\usepackage{amssymb}	
\usepackage{verbatim}
\usepackage{placeins}
\usepackage{float}
\usepackage{mathtools}
\usepackage{multirow}
\usepackage{multicol}

\newcommand{\Mstar}{\ensuremath{M_\star}}
\newcommand{\lambdar}{\ensuremath{\lambda_R}}
\newcommand{\Reff}{\ensuremath{R_{\rm e}}}
\newcommand{\ellip}{\ensuremath{\epsilon}}
\newcommand{\afour}{\ensuremath{a_4 / a}}
\newcommand{\vavg}{\ensuremath{V_{\rm avg}}}
\newcommand{\vdisp}{\ensuremath{\sigma}}
\newcommand{\hthree}{\ensuremath{h_3}}
\newcommand{\hfour}{\ensuremath{h_4}}
\newcommand{\kpc}{\ensuremath{\rm kpc}}

\newcommand{\hthreepar}{\ensuremath{\xi_3}}

\newcommand{\fzpro}{\ensuremath{f^{\rm pro}_{\rm z-tube}}}
\newcommand{\fxtube}{\ensuremath{f_{x-tube}}}
\newcommand{\agn}{\textit{AGN}}
\newcommand{\noagn}{\textit{NoAGN}}
\newcommand{\atlas}{\ensuremath{{\rm ATLAS}^{3D}}}
\newcommand{\triax}{\ensuremath{T}}

\title[Impact of AGN on stellar kinematics]{The impact of AGN on stellar kinematics and orbits in simulated massive galaxies}
\author[M. Frigo, T. Naab et al.]
{Matteo Frigo$^{1}$\thanks{E-mail: mfrigo@mpa-garching.mpg.de (MF)},
 Thorsten Naab$^{1}$, Michaela Hirschmann$^{2,3}$, Ena Choi$^{4}$, \newauthor 
Rachel S. Somerville$^{5,6}$, Davor Krajnovic$^{7}$, 
 Romeel Dav\'e$^{8,9}$ and \newauthor 
Michele Cappellari$^{10}$\\
$^{1}$Max-Planck-Institut fuer Astrophysik, Karl-Schwarzschild-Strasse 1, 85741 Garching, Germany\\
$^{2}$Institut d' Astrophysique de Paris, F-75014 Paris, France\\
$^{3}$University of Vienna, Department of Astrophysics, T\"urkenschanzstrasse 17, 1180 Vienna, Austria\\
$^{4}$Department of Astronomy, Columbia University, New York, NY 10027, USA\\
$^{5}$Center for Computational Astrophysics, Flatiron Institute, 162 5th Ave, New York, NY 10010, USA \\
$^{6}$Department of Physics and Astronomy, Rutgers University, 136 Frelinghuysen Road, Piscataway, NJ 08854, USA \\
$^{7}$Leibniz-Institut fur Astrophysik Potsdam (AIP), An der Sternwarte 16, D-14482 Potsdam, Germany \\
$^{8}$Institute for Astronomy, University of Edinburgh, Royal Observatory, Edinburgh EH9 3HJ, UK \\
$^{9}$Department of Physics and Astronomy, University of the Western Cape, Bellville, Cape Town 7535, South Africa \\
$^{10}$Sub-Department of Astrophysics, Department of Physics, University of Oxford, Denys Wilkinson Building, Keble Road, Oxford OX1 3RH, UK}

\begin{document}

\date{Accepted --. Received --; in original form --}

\pagerange{\pageref{firstpage}--\pageref{lastpage}} \pubyear{2018}

\maketitle

\label{firstpage}

\begin{abstract}

We present a series of 20 cosmological zoom simulations of the formation of massive 
galaxies with and without a model for AGN feedback. Differences in stellar population and 
kinematic properties are evaluated by constructing mock integral field unit (IFU) maps. 
The impact of the AGN is weak at high redshift when all systems are mostly fast rotating 
and disc-like. After $ z\sim 1$ the AGN simulations result in lower mass, older, less 
metal rich and slower rotating systems with less disky isophotes - in general agreement 
with observations. Two-dimensional kinematic maps of in-situ and accreted stars show 
that these differences result from reduced in-situ star formation due to AGN feedback. 
A full analysis of stellar orbits indicates that galaxies 
simulated with AGN are typically more triaxial and have higher fractions of x-tubes 
and box orbits and lower fractions of z-tubes. This trend can also be explained by
reduced late in-situ star formation. We 
introduce a global parameter, $\xi_3$, to characterise the anti-correlation between 
the third-order kinematic moment \hthree\ and the line-of-sight velocity ($v_{los}/\sigma$), 
and compare to ATLAS$^{3D}$  observations. The kinematic asymmetry parameter $\xi_3$ 
might be a useful diagnostic for large integral field surveys as it is a kinematic indicator 
for intrinsic shape and orbital content.     
\end{abstract}

\section{Introduction}
\label{Sec:Introduction} 
\begin{figure*} 
\centering
\includegraphics[width=\textwidth]{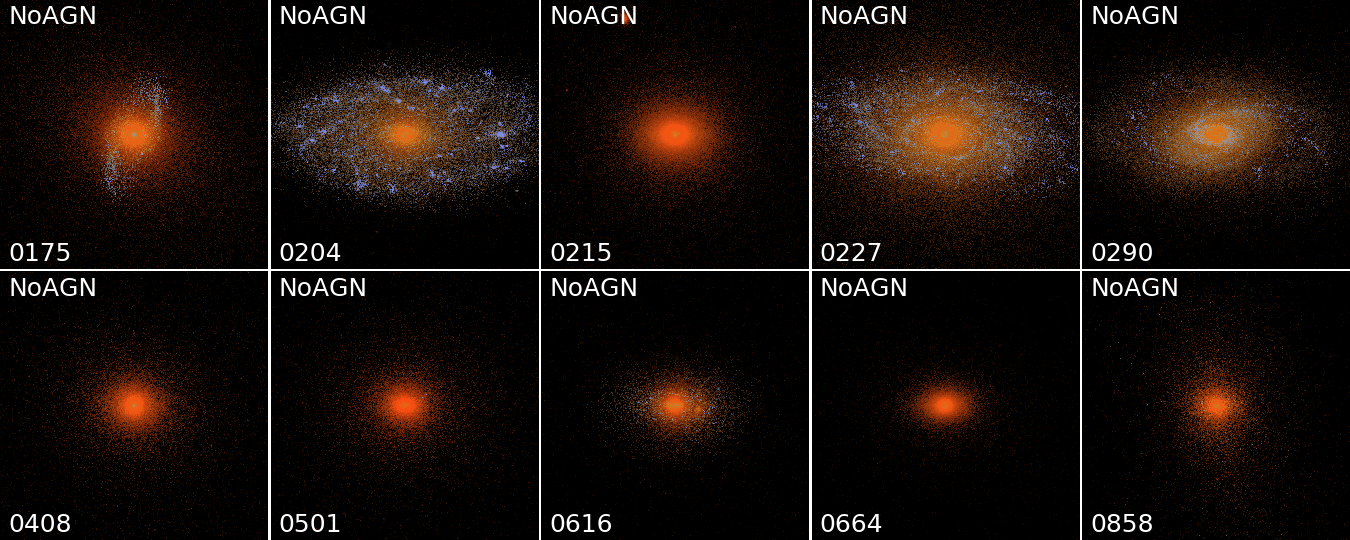} 
\includegraphics[width=\textwidth]{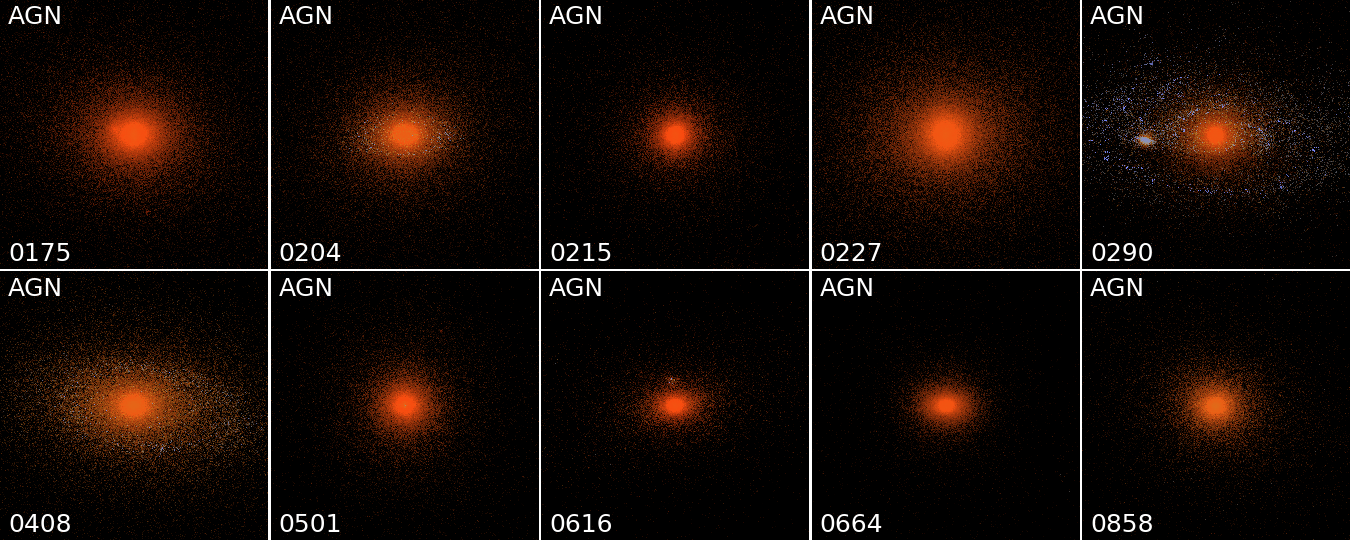}    
\caption{Mock luminosity images of our sample of simulated galaxies,
  run without (top two rows) and with (bottom two rows) AGN feedback. 
  All galaxies are viewed at an angle of 30 degrees. 
  Stars are colour-coded by V-band weighted age based on
  \citet{2003MNRAS.344.1000B}. Many of the galaxies simulated without
  AGN feedback show the presence of young stellar discs, despite
  being massive ellipticals.} 
\label{fig:starmap}
\end{figure*}

The connection between active galactic nuclei (AGN) and their host
galaxies has been subject of research for more than two decades. Soon
after the discovery of super-massive black holes (SMBH) in the centres
of early-type galaxies, correlations have been found between their mass
and galactic properties such as galactic bulge mass and
velocity dispersion \citep{dressler1989,kormendy1993,gebhardt2000}. 
This connection has been in the focus of theoretical work with the 
conclusion that the energy feedback from accreting black holes could be 
necessary to reproduce these scaling relations as well as 
the correct masses and abundances of early-type  
galaxies in cosmological simulations (see e.g. 
\citealp{2006MNRAS.365...11C, eagle, illustrisnature} 
and reviews by \citet{2013ARA&A..51..511K}, 
\citet{2015ARA&A..53...51S} and \citet{2017ARA&A..55...59N}).  
However, the impact of AGN might go beyond affecting global properties. 
The cosmological simulations of \citet{choi2015, 2017ApJ...844...31C} showed that in 
low-redshift galaxies the fraction of stars that form in-situ is much 
lower when including AGN feedback. This has strong repercussions on the 
morphological and kinematic properties of these galaxies:
in-situ formed stars tend to form orderly-rotating discs, while stars 
which are accreted from other galaxies form round dispersion-supported 
systems. Because of this connection, many studies attributed the 
difference in properties of present-day galaxies to stellar origins. 
The more massive early-type galaxies, whose stellar component 
has been for a significant part accreted, tend to have smaller angular 
momentum \citep{emsellem2011} and more complex kinematics \citep{krajnovic2011}, 
while intermediate and low-mass galaxies, which have formed most of their 
stars in-situ, are simple fast-rotating systems (see \citet{cappellari2016} 
for a review). \citet{naab2014} and \citet{roettgers2014} 
linked the present-day kinematics of simulated galaxies to the 
type of galaxy mergers they experienced during their formation: minor or 
major, and with or without gas. This picture would however be incomplete
without including AGN feedback.  
\citet{dubois2016} and \citet{penoyre2017} showed that only with
AGN feedback they were able to obtain realistic abundances of slow-rotating
systems in cosmological simulations. 
In this paper we analyse a small sample of high-resolution `zoom' simulations
(see Figure \ref{fig:starmap}) for a more in-depth look at the 
impact of AGN feedback on the kinematic and stellar-population properties of galaxies,  
but also extending the analysis to higher-order kinematics and orbital structure.
We compare our simulated galaxies to observations by mocking the images produced 
by integral field unit (IFU) spectrographs. 
These instruments collect a spectrum for each of their spatial pixels, 
so that one can observe the spatial distribution of spectrum-derived quantities, 
such as line-of-sight velocity, metallicity and age. Recently a number of 
large galaxy surveys have been performed with IFU spectrographs, such as MaNGA 
\citep{mangaoverview}, SAMI \citep{samioverview} and CALIFA \citep{califaoverview}, 
resulting in the mapping of thousands of galaxies. The MUSE spectrograph 
\citep{museoverview} also delivered detailed 2D maps of galactic 
properties (e.g., \citealp{emsellem2014}, \citealp{krajnovic2018}), including at high redshift \citep{guerou2016}. For the study of AGN 
feedback, this means that there is a huge library of data that can be used to 
look for signatures of the effect of AGNs, and in this paper we want to 
understand the nature of generic signatures through cosmological simulations. 
We do this by running each of our simulations twice, once with and once without 
our AGN feedback implementation, in order to find and analyse the differences 
between the two cases. 
In Section \ref{sec:comp} we present 
the set of cosmological simulations analysed in this paper. 
In Section \ref{sec:anal} we describe how our mock observational maps are 
created, and how the other values we present are calculated. In Section 
\ref{sec:exem} we look at the effect of AGN feedback on one exemplary simulated 
galaxy, through our mock integral field maps. In Section \ref{sec:gen} we 
analyse the full simulation sample, to get an idea of the general impact 
of AGN feedback. In Section \ref{sec:conc} we discuss and summarise our 
conclusions.

\section{Simulation details} \label{sec:comp}

\subsection{Cosmological `zoom' simulations}
In this work we analyse a set of twenty prototypical cosmological 
{\it zoom} simulations of massive 
galaxies for the impact of feedback from accreting super-massive 
black holes. Each initial condition is simulated once   
with and once without the AGN feedback model. Throughout the 
paper the two cases will be labelled as \agn\ and \noagn.
The initial conditions for the simulations were constructed
from a  $(100 \, {\rm Mpc})^3$ dark matter only simulation with
a WMAP3 cosmology \citep{wmap3}: $h = 0.72, \Omega_{\rm b}=0.044,
\Omega_{\rm dm}=0.216, \Omega_\Lambda=0.74, \sigma_{\rm 8} = 0.77,
n_{\rm s} = 0.95$ . All details on the construction of 
the zoom initial conditions are presented in \citet{oser2010,oser2012}. 
The same initial conditions were used in e.g. \citealp{naab2014} and 
\citealp{2012MNRAS.419.3200H, 2013MNRAS.436.2929H, naab2014} and \citealp{michaela2017}, 
but here we simulate at higher resolution. 
Dark matter particles have a mass of $m_p = 3.62 \cdot 10^6 \, M_\odot \, h^{-1}$ 
and gas particles initially have mass of
$m_p = 7.37 \cdot 10^5 M_\odot \, h^{-1}$. The simulations are run from
$z=43$ to $z=0$ with gravitational softening lengths of $0.2 \, \kpc$
for gas, star and black hole particles and $0.45 \, \kpc$ for
dark matter particles at the highest resolution level. \newline
The simulation code is the same as the one used in \citet{michaela2017}. We used an improved version of {\small GADGET}3
\citep{gadget2}, SPHgal, which overcomes the numerical limitations of the classic smoothed
particle hydrodynamics (SPH) implementation. All details of SPHgal are
given in \citet{chiayu2014}. We use a pressure-entropy SPH formulation,
 a Wendland $C^4$ kernel with 200 neighbouring particles, artificial
thermal conductivity and an improved modelling of artificial viscosity.

\subsubsection{Star formation and feedback}

The simulation code includes a model for the formation of stellar populations from
gas particles, representing star formation and for feedback. The stellar 
populations provide thermal and kinetic feedback as well as metals to the 
inter-stellar medium. The chemical enrichment was originally described 
by \citet{cs2005,cs2006} and later improved by \citet{aumer2013} 
and \citet{nunez2017} with an updated feedback model. Gas particles 
are stochastically converted into star particles depending on the 
density of the gas, in a way that reproduces the Kennicutt-Schmidt 
relation \citep{kennicutt1998}. To be eligible for conversion into stars, 
SPH particles need to have a temperature lower than $12,000 \, \rm K$ 
and a density higher than $1.94 \times 10^{23} \, \rm g \, cm^{-3}$. 
The probability for conversion during a time step of $\delta t$ is 
$1 - e^{-p_{\rm SF}}$, where:   
\begin{equation}
p_{\rm SF} = \epsilon_{\rm SFR} \, \sqrt{4 \, \pi \, G \, \rho} \, \delta t
\end{equation}
and $\epsilon_{\rm SFR}$ is set to $0.02$ (see
e.g. \citealp{2000MNRAS.312..859S}). The newly-created star particles are
then treated as collisionless. Each particle represents a single
stellar population assuming a \citet{kroupaimf} initial mass function, 
with a given age and the metallicity of the original gas particle. 
This stellar population then exerts
feedback to the surrounding gas. This takes the form of type Ia and II
supernovae and of winds from asymptotic giant branch (AGB)
stars. Supernovae Type II happen at a given time $\tau_{SNII} = 3\, \rm Myr$ 
after the creation of the star particles, where $\tau_{SNII}$. 
This is on the short end of the typical delay time distributions 
for supernovae type II. Supernovae type Ia and AGB winds are added 
continuously every $50 \rm Myr$ after the star particle creation. 
Each event provides momentum and thermal energy to the surrounding gas. 
The total feedback energy is given by:  
\begin{equation}
E = \frac{1}{2} \, m_{\rm ejected} \, v_{\rm out}^2,
\end{equation}
where $m_{\rm ejected} $ is the mass ejected by the stellar population and
$v_{\rm out}$ is the assumed ejecta velocity. These are determined depending 
on the mass, age and metallicity of the particle and on the type of
event. We assume $v_{\rm out} = 4500 km s^{-1}$ for SNIa and SNII and 
$v_{\rm out} = 10 km s^{-1}$ for AGB stars. The ejecta mass is taken 
from \citet{woosley1995} for SNII and from \citet{iwamoto1999} for SNIa. 
This energy and mass is then added to the surrounding gas both as thermal 
(heating) and as momentum feedback (pushing). The relative fraction depends 
on the density and distance between the supernova-undergoing stellar  
particle and the 10 neighbouring gas particles, mimicking the 
evolution of blast waves (a simplified version of the three-phase model adopted
in \citet{nunez2017}; see \citet{michaela2017}). The feedback events also 
distribute metals to the surrounding gas. Eleven elements are tracked 
for every gas and star particle (H, He, C, N, O, Ne, Mg, Si, S, Ca, and Fe), 
and their abundances are used to compute the cooling rate of the gas with 
the yields from \citet{karakas2010}, \citet{iwamoto1999} and \citet{woosley1995} 
for AGB winds and supernovae type Ia and II, respectively. All details can 
be found in \citet{aumer2013}, \citet{2014MNRAS.441.3679A}, and 
\citet{nunez2017}. 

\subsubsection{AGN feedback}
AGN feedback is represented through the model developed by \citet{choi2012} 
and used in \citet{2014MNRAS.442..440C, choi2015, 2017ApJ...844...31C}.
This model includes both a radiative and a kinetic (wind) component (see
\citealp{2017ARA&A..55...59N} for a discussion of alternative numerical 
implementations). For massive galaxies this results in efficient and 
realistic suppression of star formation, as well as good agreement with 
the observed black hole mass relations and X-ray luminosities
\citep{choi2015,eisenreich2017}. Here we summarise the most important
elements used for this study. Black holes are first seeded at the centre 
of halos exceeding a mass of $10^{11} \, M_\odot$, with an initial mass of
$M_{\mathrm{BH}} = 10^{5} \, M_\odot$. They can then grow either by merging with 
other black hole particles or by accreting
neighbouring gas particles according to a modified Bondi-Hoyle-Lyttleton 
(\citealp{1939PCPS...35..405H}, \newline \citealp{1944MNRAS.104..273B}, \citealp{1952MNRAS.112..195B})
accretion rate:  
\begin{equation}
\dot{M}_{\rm BHL} = \left< \frac{4 \, \pi \, G^2 \, M_{\rm BH}^2 \, \rho }{(c_s^2 + v^2)^{3/2}} \right> ,
\end{equation}
where $M_{\rm BH}$ is the mass of the super massive black hole, $\rho$
is the density of the gas, $v$ its relative speed and $c_s$ is its speed
of sound. The angle brackets indicate SPH kernel averaging. Of the
gas particles which could be accreted, 90\% are re-emitted as a wind parallel
to the angular momentum of the gas next to the black hole (see
\citealp{2010ApJ...722..642O}). This simulates the broad-line winds
commonly emitted by AGN \citep{dekool2001,2015ARA&A..53...51S,2017ARA&A..55...59N}. 
The remaining 10\% are accreted, increasing the mass of the black
hole particle. The model also  includes radiative feedback, in two
forms. There is an Eddington radiation pressure force, which depends
on the accretion rate and represents low energy photons providing
momentum to the gas  isotropically. We are also representing the
higher energy X-ray photons, using the formulae from
\citet{sazonov2005} for Compton scattering. This component provides
both momentum and thermal energy to the gas. 
As in \citet{michaela2017}, our simulation code differs 
from the one used in \citet{2017ApJ...844...31C} in not including 
metallicity-dependent heating, which was shown to have negligible impact\citep{2017ApJ...844...31C}.

\section{Analysis methods}\label{sec:anal}

\subsection{Voronoi-binned kinematic maps} \label{sec:mockkin}
We study our simulations by analysing a series of mock integral field unit 
(IFU) maps (e.g. \citealp{atlas3d}) of the kinematics, metallicities and ages 
of the stellar populations of the simulated galaxies. These maps are 
constructed with a Python code developed for this work, following the 
analysis presented in \citet{jesseit2007,jesseit2009,roettgers2014,naab2014}. 
The code is included in the publicly available {\small PYGAD} analysis package 
\footnote{ https://bitbucket.org/broett/pygad}. Positions and velocities of 
the simulated galaxies are centred on the densest nuclear regions using a 
shrinking sphere technique on the stellar component. In the \agn\ simulations 
we centre the galaxies on their central super-massive black hole particles, 
which we define as the most massive black hole particle within $1 \kpc$ of 
the stellar density centre. We then calculate the eigenvectors of the reduced 
inertia tensor \citep{2005ApJ...627..647B} of all stellar particles within 10 \% 
of the virial radius, and use them to align the
galaxies' principal axes with the coordinate systems, such that the x-axis is 
the long axis and the z-axis is the short axis.
To mimic seeing effects, each star particle in the simulation
is split into 60 `pseudo-particles', which keep the same velocity as the 
original particle and the positions are distributed according to a
Gaussian with $\sigma = 0.2 \, \rm kpc$ centred on the original
position of the particle (see \citealp{naab2014}). In projection, the 
pseudo-particles are mapped onto a regular two-dimensional grid, with pixel 
size $0.1 \, \rm kpc$ (at $z=0$). Adjacent bins of this grid are then joined 
so that each resulting spaxel has a similar signal-to-noise
ratio. We do this using the Voronoi tessellation method presented in
\citet{vorotess}. For the maps presented in this paper, the target
signal-to-noise level of the spaxels has been set such that 
each of them contains $25000$ pseudo-particles ($\sim 400$ regular particles).
This ensures good statistics and results in maps similar to the ones from
modern integral field surveys. The Voronoi grid is then used to
construct the plots of stellar kinematics, metallicity and age shown
in this paper. For the age and metallicity maps, the value of every spaxel is
calculated through a mass-weighted sample average. 
For the kinematic maps, we construct a histogram of the 
line-of-sight (LOS) velocity distribution of each spaxel, with 
the bin size determined by the \citet{Freedman1981} rule.
We then follow the classic 
approach of \citet{vdm1993,gerhard1993} and fit the LOS velocity 
histogram with a Gauss-Hermite function:
\begin{equation}
f ( V )  = I_0 \, e^{ - y^2 / 2 } (1+h_3 \, H_3 (y)+h_4 \, H_4 (y))
\end{equation}
where $y = (V - \vavg) / \vdisp$ and $H_3$ and $H_4$ are the Hermite 
polynomials of third and fourth order:
\begin{equation}
H_3(y) =  (2 \, \sqrt{2} \, y^3 - 3 \, \sqrt{2} \, y) / \sqrt{6}
\end{equation}
\begin{equation}
H_4(y) =  (4 \, y^4 - 12 \, y^2 + 3) / \sqrt{24}
\end{equation}
The four fitting parameters \vavg\ (average velocity), \vdisp\ 
(velocity dispersion), \hthree\ (skewness of the distribution), \hfour\ 
(kurtosis of the distribution) are the ones plotted in the four panels 
of the kinematic maps. \newline
To characterise the angular momentum of our galaxies we also employ the
$\lambda_R$ parameter \citep{2007MNRAS.379..401E}, defined as: 
\begin{equation} \label{eq:lambdar}
\lambda_R=\frac{\sum_i F_i R_i | V_i |}{\sum_i F_i R_i \sqrt{V_i^2 + \sigma_i^2}},
\end{equation}
where the sum has been carried out over the spaxels of the kinematic
maps, and $F_i$, $R_i$, $V_i$ and $\sigma_i$ are the flux, projected
radius, average LOS velocity, and velocity dispersion of each spaxel,
respectively. By limiting the sum to bins within a certain radius $R$, it
is also possible to evaluate the cumulative radial $\lambda_R$ profile for every
galaxy. The values given in the tables and kinematic maps are 
all calculated within \Reff . 

\begin{figure}
\centering
\includegraphics[width=\columnwidth]{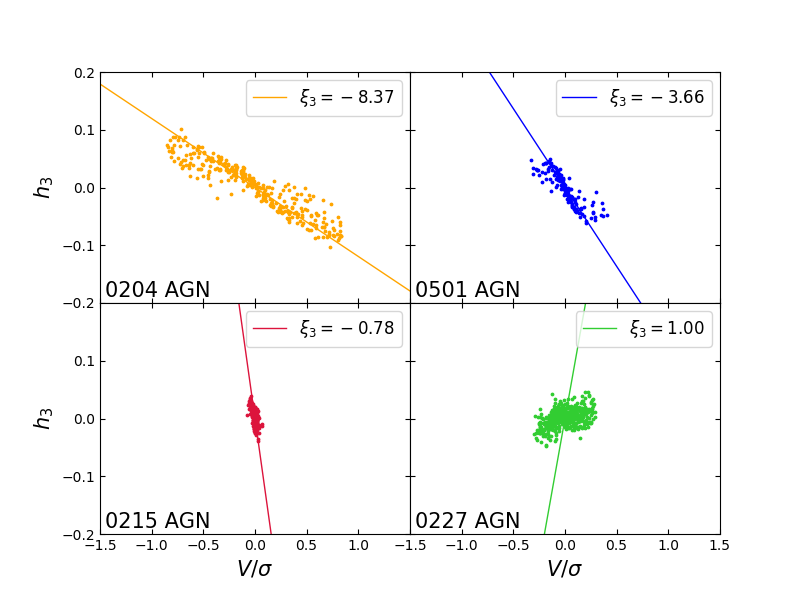}
\caption{Four simulation examples for the relation between \hthree\ and 
\vavg/\vdisp\ for the Voronoi bin within \Reff\ with the corresponding 
$\hthreepar$ values. The lines indicate 
$\hthree = (1 / \hthreepar ) \cdot \vavg / \vdisp $. A typical axisymmetric 
fast rotator (0204 \agn) is shown in the upper left plot. It has the most 
negative $\hthreepar$ value. The more complex fast rotator 0501 \agn 
(upper right panel) shows a steeper slope with a higher $\hthreepar$. 
Slow rotators like 0215 \agn\ (bottom left panel) have  $\hthreepar$ close 
to zero. Unusual non-axisymmetric rotating systems like 0227 \agn have 
a weak positive correlation between \hthree\ and \vavg / \vdisp\, resulting 
in a slightly positive value of \hthreepar (bottom right panel).}
\label{fig:h3example}
\end{figure}

\subsection{Higher-order kinematics} \label{sec:h3def}
The higher-order moments of the LOS velocity distribution, \hthree\ and \hfour, can
provide additional information on the orbital structure of our galaxies.
In rotating systems the \hthree\ parameter has been observed to be
anti-correlated to the average LOS velocity \vavg, or more specifically to the 
\vavg/\vdisp\ ratio \citep{gerhard1993, krajnovic2011, veale2017, samih3}. 
This anti-correlation indicates that the LOS velocity distributions typically 
have a steep leading wing and a broad trailing wing. Simple axisymmetric rotating 
stellar systems show this property due to projection effects - stars are typically 
on circular orbits and those with lower LOS velocities projected into each spaxel 
produce a broad trailing wing. The slope of this anti-correlation is then about 
$\sim 0.1$ \citep{bender1994}. However, if the galaxy is more complex, i.e not 
axisymmetric, it can also contain stars orbiting around different axes or radial 
orbits. This can make the trailing wing broader, as these stars have lower LOS 
velocities. The slope of the anti-correlation would then be steeper, and in some 
slow-rotating galaxies it can become extremely steep (see e.g. \citealp{samih3}). 
If he group of rotating stars becomes sub-dominant the correlation between \hthree\ 
and  \vavg\ / \vdisp\  can change sign and become positive.
Here the few fast rotating stars create a broad leading wing in the LOS velocity 
distribution \citet{2006MNRAS.372..839N,hoffman2009,roettgers2014}. This unusual 
property is typically seen in simulated gas poor mergers \citet{2001ApJ...555L..91N,naab2014}.

We characterise this variety of behaviours with a global parameter indicating 
the slope of the relation between \hthree\ and  \vavg\ / \vdisp\ for all spaxels 
of one galaxy. This definition is inspired by the finding in \citet{naab2014} 
that different slopes indicate varying formation histories and by the improved 
empirical classifications of the SAMI and MASSIVE galaxy surveys \citep{samih3,veale2017}. 
We define \hthreepar\ as:
\begin{equation}\label{eq:hthreepar}
\hthreepar = \frac{<\hthree \cdot \vavg/\vdisp>}{<{h_3}^2>}= 
\frac{\sum_i F_i \, h_{3,i} \cdot (V_i/\sigma_i)}{\sum_i F_i \, {h_{3,i}}^2},
\end{equation}
where the sum is calculated over each spaxel out to \Reff\ from the centre.
When \hthree\ and \vavg/\vdisp\ are correlated, this parameter 
estimates the inverse of the slope of the correlation to reasonable 
accuracy with a simple fraction of weighted sums; negative values 
indicate a negative correlation, while positive values indicate a positive one. 
This can be seen by assuming $<h_3> = 0$ and $<\vavg/\vdisp> = 0$ and rewriting 
the definition of the parameter as:
\begin{equation}\label{eq:hthreepearson}
\hthreepar = \rho_{V/\vdisp , \hthree} \, \frac{\sigma_{V/\vdisp}}{\sigma_{\hthree}},
\end{equation}
where $\rho_{V/\vdisp , \hthree}$ is the \citet{1895RSPS...58..240P} correlation 
coefficient of \vavg/\vdisp\ and \hthree , and $\sigma_{V/\vdisp}$ 
and $\sigma_{\hthree}$ are the dispersion values of the two parameters. 
If \hthree\ and \vavg/\vdisp\ are linearly correlated then $\rho = \pm -1$, 
and \hthreepar\ becomes exactly the slope of the correlation.
Figure \ref{fig:h3example} shows an example of the  
\hthree\ - \vavg\ / \vdisp\ spaxel values within \Reff\ for four simulated 
galaxies with different LOS velocity distribution properties.  The lines 
indicate the simple slope given by $\hthree = (1 / \hthreepar ) \cdot \vavg / \vdisp$. 
Purely rotating systems are expected to have $\hthreepar \sim -10$ or lower, 
while rotating systems with
non-negligible fractions of different orbit types are expected to lie in the 
$-3 <\hthreepar < -6$ range. When there is no correlation, or when
the slope is almost vertical (both of which are observed in slow-rotating 
galaxies), the value of \hthreepar\ comes close to zero.
Additionally, the dependence of \hthreepar\ on inclination seems to be weaker than other kinematic global parameters, 
making it potentially a good way of distinguishing different types of 
galaxies. In Section \ref{sec:hthreepar} we investigate inclination 
effects and show how this parameter correlates with other galaxy properties. 
\citet{samih3} have used best fitting elliptical Gaussians with a 
maximum log-likelihood approach to characterise the slope of the relation, 
which is slightly more complicated than our procedure. \citet{veale2017} 
perform linear least square fits to calculate the slopes directly. 
Using the inverse of the slopes highlights the difference between slow 
rotators and the slow rotators get values around zero.  
The \hfour\ parameter is known to relate to orbit anisotropy 
\citep{vdm1993, gerhard1993,thomas2007}, with negative values indicating 
the dominance of tangential orbits and positive values
corresponding to radial orbits. We do not further analyse  \hfour\ other 
than showing the projected maps. 

\begin{figure}
\centering
\includegraphics[width=\columnwidth]{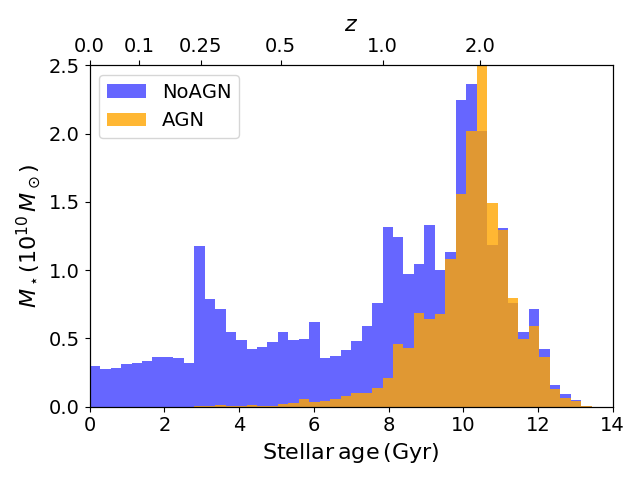}
\caption{Age distribution of star particles in the case study galaxy (0227) 
for the run with AGN feedback (orange) and the one without (blue). 
The top x-axis shows the corresponding redshift at which the stars have formed. 
Star formation proceeded at a similar rate up $z=2$. Then it is rapidly 
terminated in the presence   of AGN feedback. Without AGN feedback star 
formation continues all the way to $z=0$.}
\label{fig:sfrhist}
\end{figure}

\subsection{Isophotal shape and triaxiality}\label{sec:photo}
Our analysis involves the calculation of photometric
quantities, such as the ellipticity \ellip\ and the isophotal shape parameter \afour. 
The ellipticity values are calculated by fitting the galaxy isophotes 
with ellipses. The isophotes are constructed as lines with constant stellar 
surface mass density. For each galaxy we use 10 isophotes between $0.25 \Reff$ and $\Reff$ 
and average their ellipticity values to obtain \ellip\ . 
The \afour\ parameter represents the deviation of the shape of the isophotes of the
galaxy from a perfect ellipse. It is used to discriminate between galaxies
with `boxy' or `disky' isophotes \citep{1985MNRAS.216..429L, 1987A&A...177...71B}. We calculate it by applying a Fourier
expansion to the deviation of the actual isophotes from their best-fitting ellipse: 
\begin{equation}
\delta \, R(\theta) =  R(\theta)-R_{ell}(\theta) = \sum_n (a_n \, \cos (n \, \theta) + b_n \, \sin (n \, \theta)),
\end{equation}
where $\theta$ is the azimuthal angle \citep{1987MNRAS.226..747J}. The first, second and third
order coefficients are negligible if the ellipse is centred correctly
and has the correct ellipticity and orientation angle. The fourth
order coefficient $a_4$, normalised to the zeroth coefficient $a$,
represents the deviation of the isophote from a pure
ellipse. A positive value of \afour\ means that there is an excess of light along
the major axis of the ellipse, causing the real isophote to be more
`disky'. A negative value instead means that the shape of the real
isophote is more `boxy' (see e.g. \citealp{1999ApJ...523L.133N,2000MNRAS.312..859S}). \newline
We also look at the 3D shape of our galaxies by computing the triaxiality parameter:
\begin{equation}
T=\frac{1-(b/a)^2}{1-(c/a)^2},
\end{equation}
where $b/a$ and $c/a$ are the ratios between the main axes. We calculated the 
axis ratios through the reduced inertia tensor \citep{2005ApJ...627..647B} of
all particles within the effective radius \Reff:
\begin{equation}
\tilde{I}_{i,j}=\sum_{{\rm Particles} \, k} m_k \frac{r_{k,i} \, r_{k,j}}{r_k^2},
\end{equation}
where $m_k$ and $\vec{r}_k$ are the masses and positions of the particles.
The square roots $\tilde{a} > \tilde{b} > \tilde{c}$ of 
the eigenvalues of this tensor are related to the real axis ratios by:
\begin{equation}
\tilde{b}/\tilde{a}=(b/a)^{\sqrt 3} \; \; \; {\rm and} \; \; \; 
\tilde{c}/\tilde{a}=(c/a)^{\sqrt 3}.
\end{equation}
When $T=0$ the galaxy is perfectly oblate, while when $T=1$ the galaxy 
is perfectly prolate.

\begin{figure*}
\centering
\includegraphics[width=\textwidth,height=\textheight,keepaspectratio]{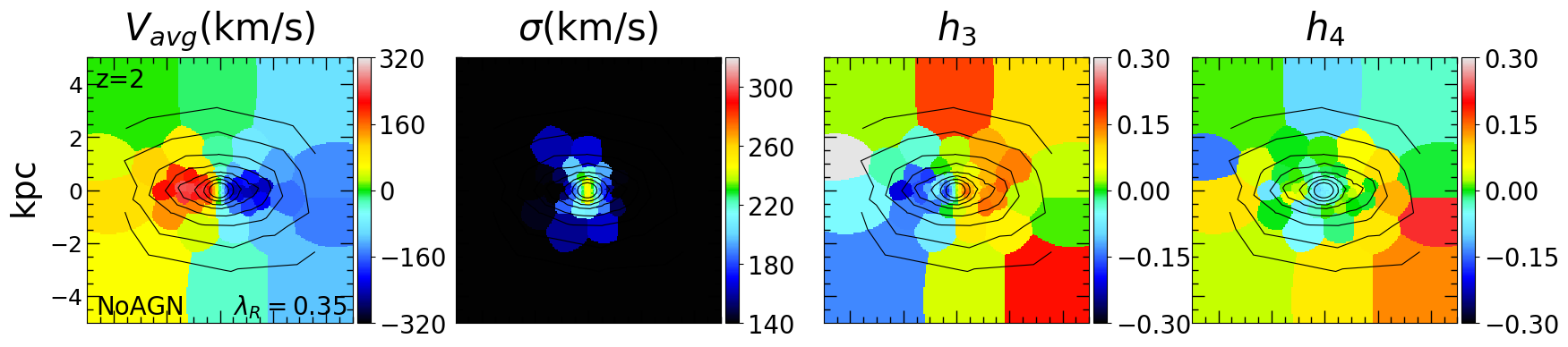}
\includegraphics[width=\textwidth,height=\textheight,keepaspectratio]{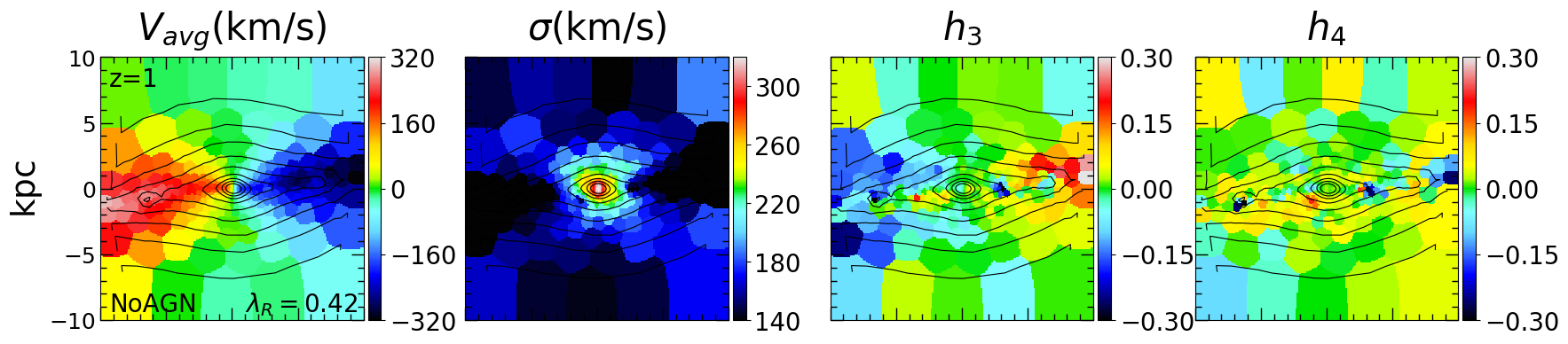}
\includegraphics[width=\textwidth,height=\textheight,keepaspectratio]{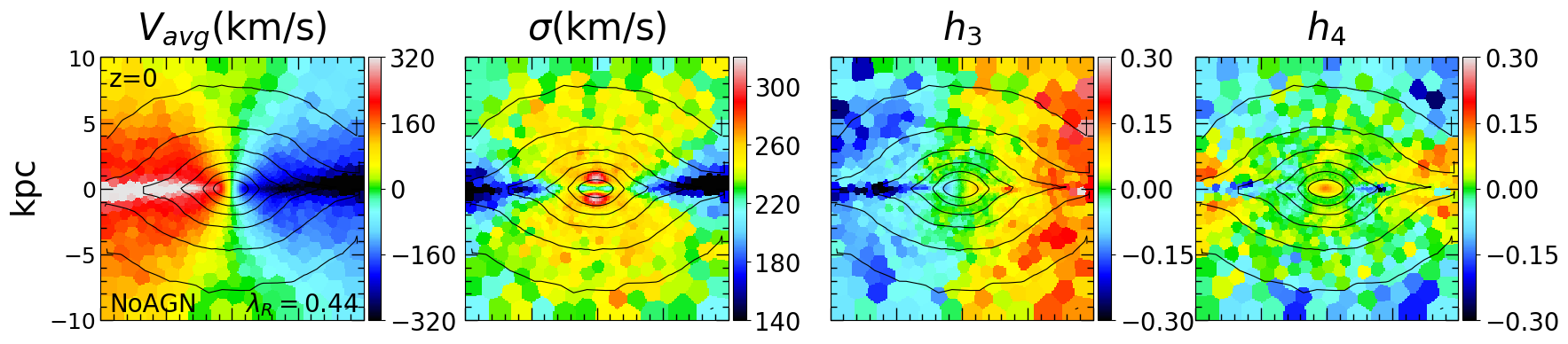}  
\caption{Edge-on two-dimensional line-of-sight stellar kinematics
  (\vavg-\vdisp-\hthree-\hfour\, from left to right) of galaxy 0227
  simulated without AGN feedback at $z=2$, $z=1$, and $z=0$ (from top
  to bottom). The maps shows typical features of systems with a
  disc-like component: high LOS velocity in the mid-plane, dumb-bell
  shaped velocity dispersion with a suppression in the mid plane disc
  region, anti-correlation of line-of-sight velocity and \hthree\, 
  negative \hfour\ along the disc. These features become strongest at
  $z=0$, when the disc is most prominent and can clearly be seen in
  the surface density contours (black lines).} 
\label{fig:kinnoagn}
\end{figure*}

\subsection{Orbit analysis} \label{sec:orbitan}

We analyse the orbital composition of each of our 
simulated galaxies following the approach of \citet{2005MNRAS.360.1185J} 
and \citet{roettgers2014}. This procedure starts by freezing the potential 
of the simulated galaxy at $z=0$ and representing it analytically using the  
self-consistent field method \citep{ho1992}: the density and potential 
are expressed as a sum of bi-orthogonal basis functions, which satisfy 
the Poisson equation. There are multiple such density-potential pairs. 
We used the one from \citet{ho1992}, in which the zeroeth-order 
element is the \citet{hernquist1990} profile:

\begin{equation}
\rho_{000} = \frac{M}{2 \, \pi \, a^3} \frac{1}{\frac{r}{a} (1 + \frac{r}{a})^3}
\end{equation}
\begin{equation}
\Phi_{000} = - \frac{G \, M}{r + a},
\end{equation}

where $a$ is the scale parameter of the Hernquist profile. Higher order
terms then account for both radial and angular deviations. We then 
integrate the orbits of each stellar particle within this fixed   
analytical potential for about 50 orbital periods. This is enough for 
identifying the orbit type, but not so much that quasi-regular orbits
diverge from regular phase-space regions forcing us to classify them
as irregular. The orbit classification itself is then done using the
code by \citet{carpintero1998}, which distinguishes different orbit 
families by looking at the resonances between their frequencies along 
different axes. In this paper we consider 4 main families 
of orbits: z-tubes (orbits that rotate around the z axis), 
x-tubes (orbits that rotate around the x-axis), box orbits ($\pi$ boxes 
and boxlets), and irregular orbits. 
In addition to these, we computed the fraction of prograde z-tube orbits 
$f^{pro}_{z-tube}$, by only selecting z-tubes with angular momentum 
along the z-axis of the same sign as the overall galaxy.

\begin{figure*}
\centering  
\includegraphics[width=\textwidth,height=\textheight,keepaspectratio]{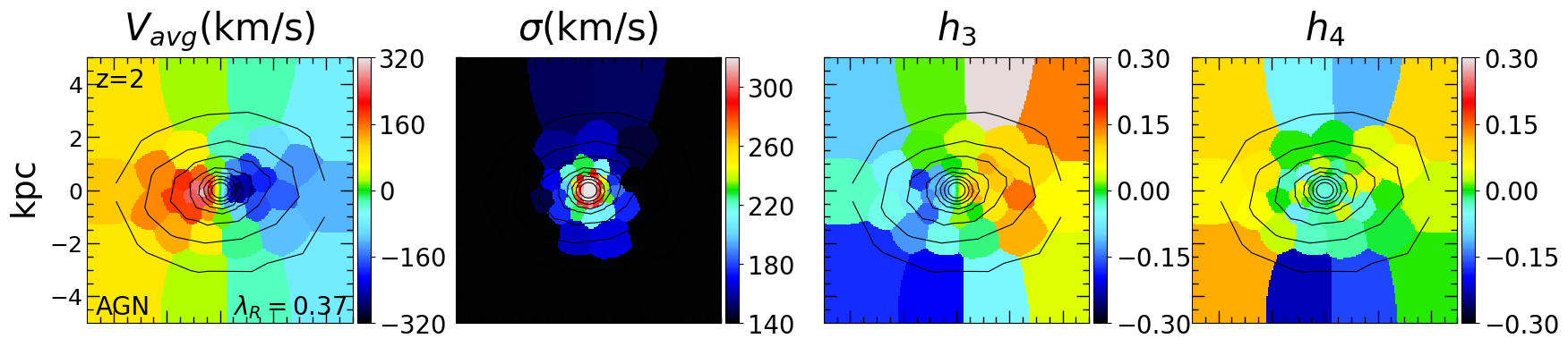}
\includegraphics[width=\textwidth,height=\textheight,keepaspectratio]{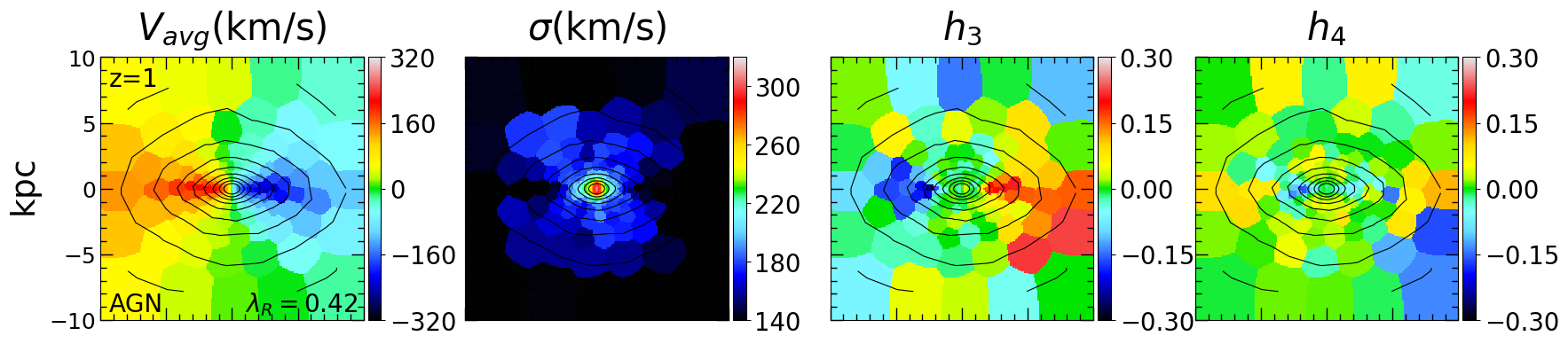}  
\includegraphics[width=\textwidth,height=\textheight,keepaspectratio]{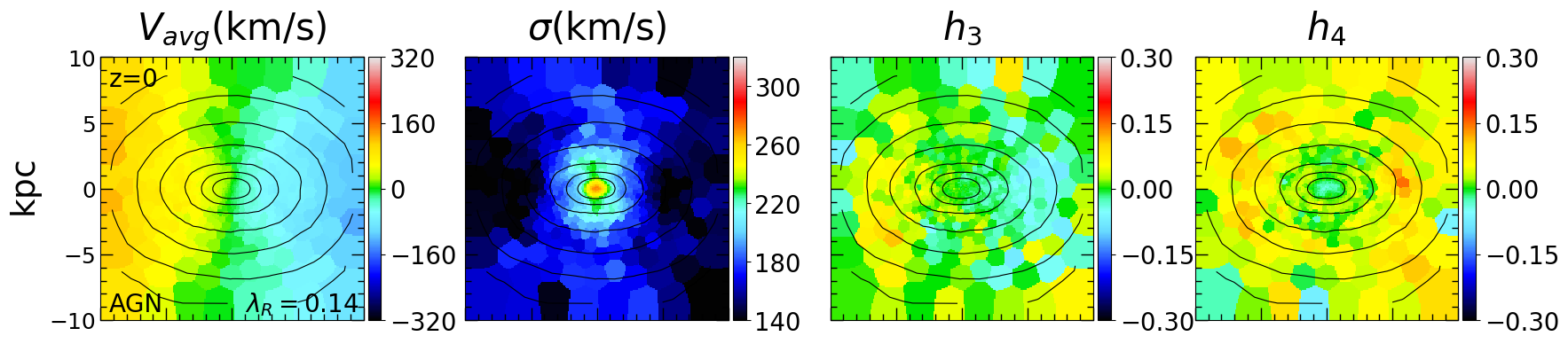}     
\caption{Same as Fig. \ref{fig:kinnoagn}  for galaxy 0227 simulated
  including AGN feedback. The kinematics is qualitatively similar to the case
  without AGN at $z=2$ and $z=1$. By $z=0$ however the strong
  rotational signatures are gone, and the galaxy looks more like a
  typical slow-rotator without kinematic disc signatures: low average
  LOS velocity, high velocity dispersion, no
  \hthree\ anti-correlation signal, positive \hfour.}  
\label{fig:kinagn}
\end{figure*}

\section[]{A typical galaxy simulated with and without AGN feedback}\label{sec:exem}

Our study involves a small sample of 20 massive galaxies. As a test case, 
in this section, we first discuss the formation history, global galaxy
properties, stellar kinematics, stellar age and metallicity, morphology and
redshift evolution for one prototypical galaxy.  Simulating this
initial condition with and without AGN feedback allows us to investigate the impact
of AGN feedback on the final properties of the galaxy.

\subsection{Formation history and global properties} \label{sec:formhist}
Galaxy 0227 is an
early-type galaxy, with an effective radius of $4.0 \, {\rm kpc}$
and a stellar mass of $2 \cdot 10^{11} \, M_\odot$ in the \agn\ 
case and $5 \cdot 10^{11} \, M_\odot$ in the \noagn\ case. Its 
formation history is characterised by a major merger at redshift 
$z \sim 0.25$, with mass ratio of $1:1.7$ and $1:1.2$ in the 
\noagn\ and \agn\ cases. The presence of AGN has a strong 
influence on the evolution after the merger. 
Figure \ref{fig:starmap} shows a mock 
V-band image of this galaxy with and without AGN feedback. In the 
absence of AGN feedback (left panel) the galaxy is still forming 
new stars in an extended disc. Instead, in the case with AGN feedback 
(right panel) the system is spheroidal with a very old stellar population.

Figure \ref{fig:sfrhist} shows the age distribution of stars in 
galaxy 0227 simulated with and without AGN feedback. The oldest stars
(age $>$ 10 Gyr) have very similar age distributions, 
with the bulk forming around $z \sim 2$. Towards lower redshifts, star formation gets 
quenched in the \agn\ case; a behaviour found in all our simulations. 
While in the \agn\ case not many stars form after $z \sim 1$,
in the \noagn\ case star formation continues throughout the simulation, 
including a starburst at $z \sim 0.25$ during the major merger.

\subsection{LOS kinematics} \label{sec:loskin}
In order to identify features in the stellar kinematics 
originating from the
impact of AGN feedback, we construct two-dimensional maps visualising kinematic properties, as detailed
in Sec. \ref{sec:mockkin}. Specifically we show the stellar line-of-sight velocity,
dispersion, and the higher order moments \hthree\ and \hfour\ in
Figs. \ref{fig:kinnoagn} and \ref{fig:kinagn} for galaxy 0227 
without and with AGN feedback at $z=2$, $z=1$, and $z=0$. 
Initially (at $z=2$ and $z=1$) there are only moderate differences between the
\agn\ and \noagn\ simulations. The \agn\ and \noagn\ galaxies (in brackets) have similar 
stellar masses of $M_* = 0.59*10^{11} \,M_\odot$ (  $M_* = 0.54*10^{11} \,M_\odot$ ) 
at $z=2$, while at $z=1$ they are $M_* \sim 1.16*10^{11}  \, M_\odot$ ( $M_* \sim 1.97*10^{11}  \, M_\odot$ ). 
The effective radii are $ \sim 0.18 \,  \rm kpc$ ( $ \sim 0.35 \,  \rm kpc$ )
at $z=2$ and $0.95 \rm kpc$ ($1.53 \rm kpc$) at $z=1$. 
Down to $z=1$, the galaxies are supported by rotation. The average stellar line-of-sight
velocities reach values of $ \sim 200  \, {\rm km/s}$, and the velocity
dispersion values around $300  \, {\rm km/s}$. The velocity increases 
only slightly from $z=2$ to $z=1$, but the rotating component
becomes more extended for both cases. The \hthree\ parameter is
anti-correlated with the LOS velocity - a typical signature for axisymmetric 
rotating systems \citep{krajnovic2011,naab2014}. 
The origin of this effect is explained in detail in Section \ref{sec:h3def}, 
as well as in
\citet{2001ApJ...555L..91N,2006MNRAS.372..839N,roettgers2014,naab2014} 
in the context of idealised models, merger simulations and cosmological
simulations.
At redshift $z=0$, the situation is markedly different. In the \noagn\
case the rotation signatures are significantly enhanced. The
LOS velocities reach up to $320 \, {\rm km/s}$ in an
extended disc. The velocity dispersion map shows a dumbbell feature
with reduced velocity dispersion in the mid plane, which is a signature of 
an edge-on rotation-supported disc embedded in a dispersion-supported 
spheroidal component. This can be seen by the isophotes (see
Sec. \ref{sec:angiso}). The LOS velocity distribution is asymmetric with
anti-correlated \hthree\ values. The \hfour\ map shows characteristic
features of disc rotation (bottom right panel of
Fig. \ref{fig:kinnoagn}). In the central kpc region, \hfour\ is
positive, indicating a more peaked Gaussian LOS velocity distribution 
with more extended wings towards lower and higher than the systemic 
velocity as individual pixels cover significant fractions of the stars' 
orbits. At larger radii (in the mid plane), \hfour\ becomes negative 
indicating coherent rotation with very weak tails towards high and low 
velocities. As \hfour\ is known to roughly correlate with the velocity 
anisotropy \citep{gerhard1993,thomas2007}, a negative \hfour\ indicates 
that tangentially biased orbits are dominating, which is to be expected 
in a rotating disc. Kinematic maps of this kind are regularly found in
observational surveys like $ATLAS^{3D}$ \citep{atlasoverview}, 
CALIFA \citep{califaoverview}, or SAMI \citep{samioverview}. 
They are, however, more common for less massive galaxies. It is very unlikely to
observe an elliptical galaxy of this high mass with such a prominent
fast-rotating disc. 
The kinematic galaxy properties are very different in the \agn\ case
(Fig. \ref{fig:kinagn}). By $z=0$, there are no signatures of a prominent rotating
stellar disc, as the AGN feedback prevents
further gas accretion and in-situ disc formation (see e.g. \citealp{2018ApJ...860...14B}). 
The galaxy is slowly rotating at $ \sim 80 \,{\rm km/s}$ and dispersion 
dominated, with only weak features in the
higher-order moments. Interestingly, \hthree\ is positively correlated
with \vavg\ in the central part of the galaxy. This is rare for observed 
galaxies, but relatively common in the simulated remnants of gas poor mergers (see
\citealp{2001ApJ...555L..91N,2006MNRAS.372..839N,roettgers2014}). This positive correlation must originate from a particular orbital distribution, which will be analysed in Section \ref{sec:orbit0227}.
Also 
a core with negative \hfour\ is still visible. Values for \hfour\ are
positive in most of the map indicating radially-biased orbits. All of
the above features are typical properties of massive early-type galaxies. 

\subsection{Age and metallicity distribution at z=0} \label{sec:agez}
\begin{figure}  
\centering
\includegraphics[width=\columnwidth]{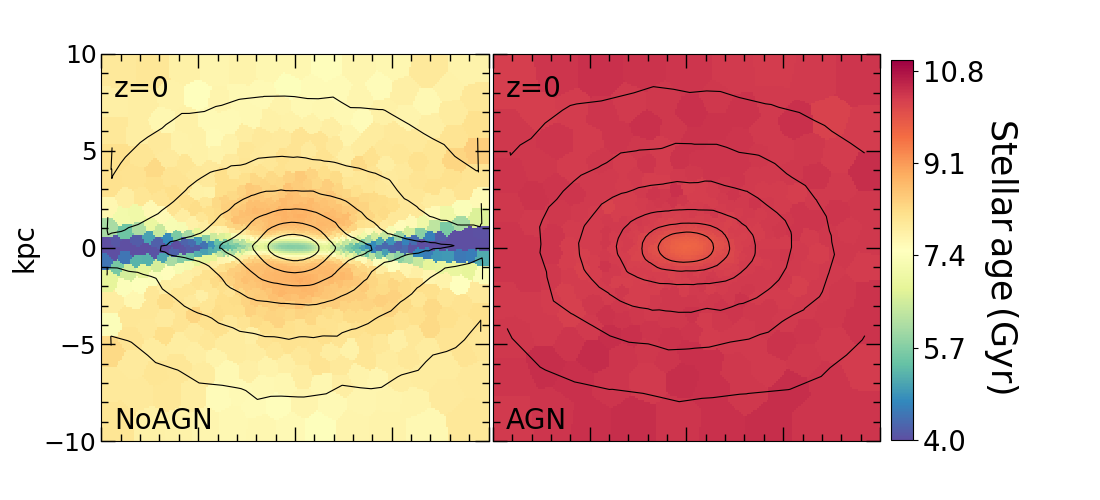}
\includegraphics[width=\columnwidth]{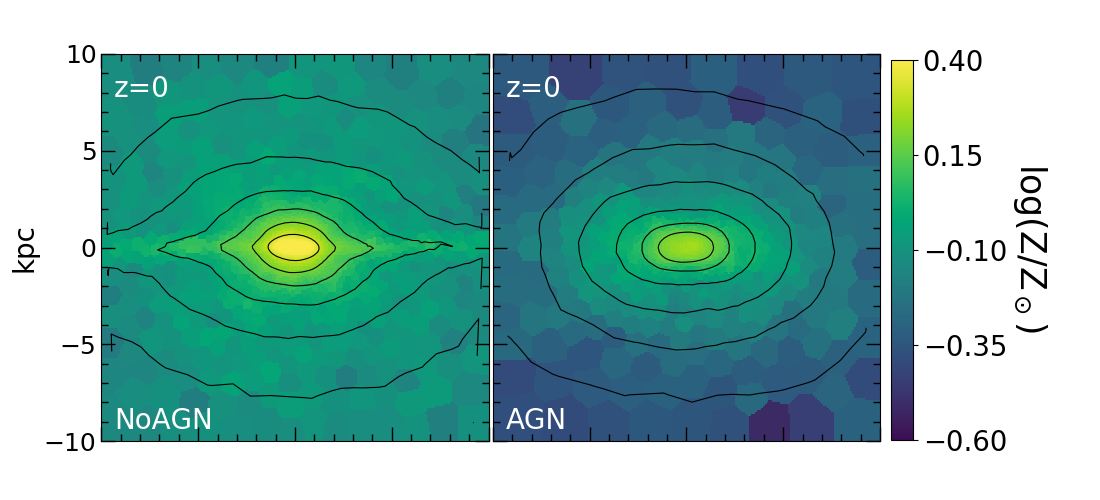}
\caption{Voronoi binned maps of the (mass-weighted) average stellar
  age (top) and metallicity (bottom) for our case-study galaxy (0227), 
  in the \noagn\ (left) and \agn\ (right) AGN cases, at $z=0$. 
  Without AGN feedback the higher star-formation
  rate at low redshift produces an overall much younger system,
  especially in the midplane, where a young stellar disc forms.
  Higher star-formation rate also result in high metallicities. 
  With AGN feedback (right panels) the galaxy is instead very old and 
  the metallicities compare well with observed early-type galaxies.} 
\label{fig:agemap}
\end{figure}

Figure \ref{fig:agemap} shows a comparison of the projected stellar
age (top panels) and metallicity (bottom panels) distributions for the \noagn\ (left column)
and \agn\ (right column) simulation at $z=0$. At low redshifts the
properties of the systems differ the most. In the \noagn\ case there is a
distinct young $\lesssim 4 \, {\rm Gyr}$ stellar disc embedded in an older 
$7 - 9 \, {\rm Gyr}$ stellar bulge. A moderate positive age gradient towards younger 
ages away from the centre is visible. The disc appears as a flattened metal 
enriched region in the mid plane, pretty much following the isophotes. 
These features indicate ongoing disc-like star formation and metal enrichment 
since $z=1$. This is also consistent with the stellar age distribution in Fig. \ref{fig:sfrhist}.       
In the \agn\ case (right panels of Fig. \ref{fig:agemap}) the stellar
population is older ($\sim 10\, {\rm Gyr}$, see also
Fig. \ref{fig:sfrhist}), less metal enriched - due to less ongoing star formation - 
with a shallower metallicity gradient. There is a mild
positive age gradient with younger ages in the centre caused by
residual nuclear star formation. The origin of age and metallicity
gradients will not be discussed further in this paper 
(see e.g. \citealp{michaela2015, 2016MNRAS.458.2371R}).

\subsection{Redshift evolution of kinematic and photometric properties}
\label{sec:angiso}
\begin{figure}
\centering
\includegraphics[width=\columnwidth]{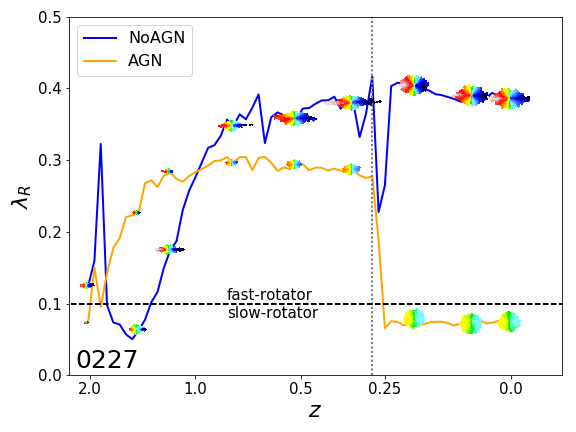}
\caption{Evolution of \lambdar\ for galaxy 0227, in the \noagn\ and \agn\ 
  cases. The values are indicated by inserted velocity
  maps out to the effective radius (isophote) of the galaxies. A
  major merger at $z \sim 0.25$ (vertical dashed line) strongly reduces the angular momentum of both systems. 
  The \noagn\ galaxy is less affected and can quickly regain angular momentum 
  due to gas accretion and star formation. The \agn\ galaxy is instead unable 
  to form new stars and remains a slow rotator.} 
\label{fig:lambdakinz}
\end{figure}

\begin{figure}
 \includegraphics[width=\columnwidth]{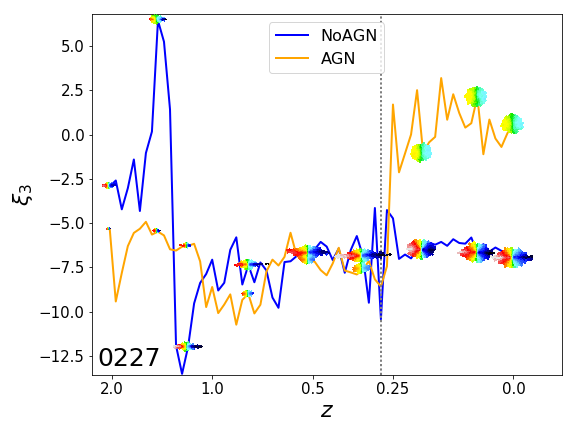}
 \caption{Evolution of \hthreepar\ for galaxy 0227, in the \noagn\ and \agn\ 
  cases. The values are indicated by small velocity
  maps out to the effective radius (isophote) of the galaxies. 
  Up to $z= 0.25$ the value of \hthreepar\ is constant for both simulations and has a value 
  as expected for a rotating system. 
  However, after a major merger at $z \sim 0.25$ (vertical dashed line), the value for the \agn\ galaxy shifts towards zero and 
  mildly positive values. This indicates that the galaxy lost its rotational support.} 
 \label{fig:h3ev}
\end{figure}

In this subsection we look at the evolution of three global 
parameters, \lambdar, \hthreepar\ and \afour, through the whole formation 
history of our case-study simulation. 
We first use \lambdar\ (Eq. \ref{eq:lambdar}) to quantify the redshift 
evolution of angular momentum in the \agn\ and \noagn\ cases. Figure 
\ref{fig:lambdakinz} shows the redshift evolution of \lambdar\ from $z=2$ to $z=0$. 
After a tumultuous phase at 
high redshift caused by mergers, at $z=1$ \lambdar\ settles at around 
$0.3$-$0.4$ in both cases. At $z=0.25$ the 
angular momentum drops because of the major merger described in Section
\ref{sec:formhist}; the vertical dashed line marks the beginning of 
this merger. The subsequent evolution diverges for the two cases. 
In the \noagn\ simulation the system is more gas rich, and thus
loses less angular momentum and even regains
it after the merger. This is a typical feature of gas rich mergers and
follow-up gas accretion (see review by \citet{2017ARA&A..55...59N}). In
the \agn\ case the system is already gas poor, without
significant star formation before the merger (see
Fig. \ref{fig:sfrhist}). The merger then reduces the
angular momentum significantly. Qualitatively this process for gas
poor mergers is discussed in detail in \citet{naab2014}. 
By $z=0$ the two systems have very different rotation properties  
with a \lambdar\ value typical of fast rotators in the \noagn\ case 
and a slow rotator value in the \agn\ case. This impact of AGN feedback 
on the rotation properties of massive galaxies has already been reported 
by \citet{2013MNRAS.433.3297D, 2014MNRAS.443.1500M} and \citet{dubois2016} for 
cosmological {\small RAMSES} adaptive mesh refinement simulations with 
different AGN feedback models. We therefore assume this to be a generic 
feature of AGN feedback. \newline
The major merger also affects the higher-order kinematic features. 
We quantify them using the parameter \hthreepar\ defined in Eq. \ref{eq:hthreepar} 
and plot it as a function of redshift, as shown in Figure \ref{fig:h3ev}. 
From $z=1$ to $z=0.25$ the two simulations show again the same behaviour, 
with the same degree of anti-correlation between \hthree\ and 
\vavg\ / \vdisp: $\hthreepar \sim -7.5$ in both cases. As discussed in section \ref{sec:h3def},   
this value is typical for a system dominated by tangential orbits, but higher 
than the one expected from a purely rotational system ($-10$). This indicates that 
a small amount of other orbit types contributes to skew the LOS velocity 
distribution. The major merger at $z=0.25$ again makes the two cases diverge. 
In the \noagn\ case the overall \hthreepar\ value stays the same. 
In the \agn\ case instead \hthreepar\ drops to $0$ and the orbital 
structure of the system is more dispersion-supported - the correlation 
between \hthree\ and \vavg / \vdisp\ becomes weaker. The sign of \hthreepar\ 
oscillates a bit, but then settles to a weakly positive value, meaning that 
\hthree\ has the same sign as \vavg\ as already pointed out. 

\begin{figure}
\centering
\includegraphics[width=\columnwidth]{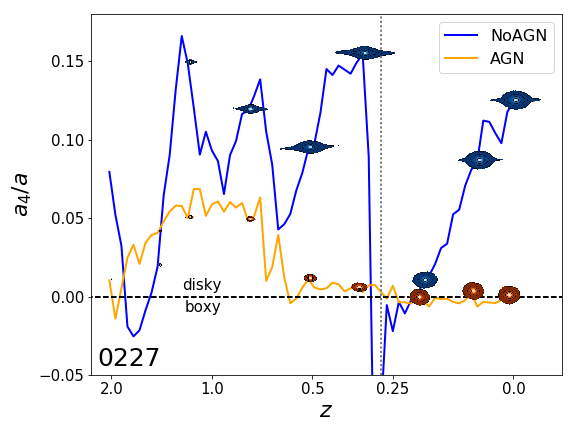}
\caption{Evolution of the isophotal shape of galaxy 0227, in the \agn\ 
  and \noagn\ case, quantified by \afour. The markers are surface
  brightness maps cut along the effective isophote. The red line
  represents perfectly elliptical isophotes. `Boxy' galaxies have
  negative, `disky' galaxies have positive \afour\ values. Without AGN
  feedback the formation of a prominent disc results in disky 
  isophotes at all times, despite the major merger at 
  $z \sim 0.25$ (vertical dashed line). The \agn\ galaxy instead loses its   
  diskyness after the merger because further star formation is suppressed by the AGN.} 
\label{fig:a4akinz}
\end{figure}

We investigate the evolution of the isophotal shape parameter \afour, 
obtained by fitting the galactic isophotes at every snapshot (see subsection 
\ref{sec:photo}), with the galaxy seen edge-on. An example of these isophotes 
can be seen in the black lines of Figs. 
\ref{fig:kinnoagn} and \ref{fig:kinagn}. We show the evolution of \afour\ 
since $z=2$ in Fig. \ref{fig:a4akinz}.  
Unlike in the previous cases, the \agn\ and \noagn\ cases are already 
different at $z=1$. The \noagn\ case has systematically higher values 
of \afour\ - more disky isophotes. This difference would however not be as 
pronounced if the galaxy was not seen from an edge-on perspective.
The value scatters due to minor mergers but 
drops to negative values after the major merger at $z=0.25$. This is the 
common feature of major 
mergers destroying previously existing disc structures (see
\citealp{1999ApJ...523L.133N,2003ApJ...597..893N}). 
Subsequently a new stellar disc forms and the \afour\ value becomes 
strongly positive again. 
In the \agn\ case the galaxy already lost 
its diskyness at high redshift, because of the suppressed inflow of 
high-angular-momentum star-forming gas, and it keeps its elliptical or mildly boxy 
isophotes to $z=0$. 
The effect of mergers and AGN feedback on the isophotal 
shape points in the same direction as the effect on \lambdar\ and \hthreepar. 

\begin{figure*}
\centering
\includegraphics[width=\textwidth,height=\textheight,keepaspectratio]{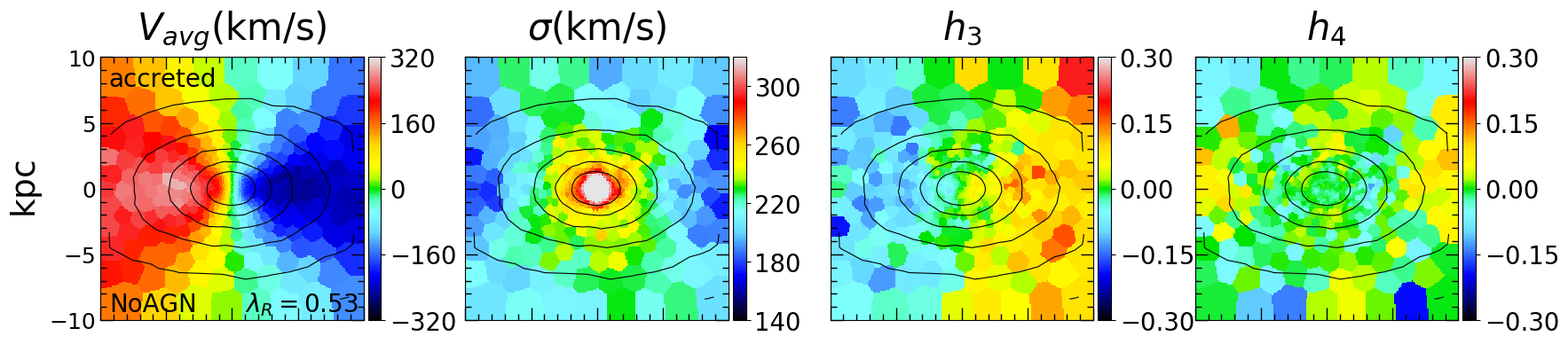}
\includegraphics[width=\textwidth,height=\textheight,keepaspectratio]{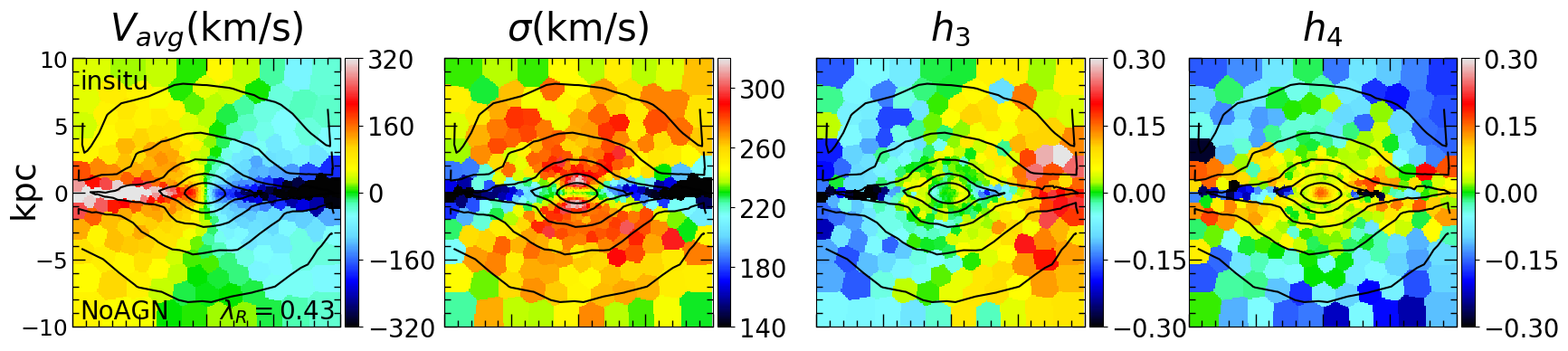} 
\caption{Stellar kinematics of galaxy 0227 (NoAGN) separated into its 
  accreted component (above) and its in-situ formed one (below). 
  The overall in-situ fraction is 50\% . The two
  components have strikingly different kinematics. 
  The accreted component is mainly pressure-supported, but also rotates fast.
  The in-situ component shows two distinct features: a fast-rotating disc 
  in the midplane with low velocity dispersion, and a slow-rotating bulge with 
  very high velocity dispersion. The disc feature formed after a recent major
  merger, while the surrounding bulge is older, and its originally rotational 
  orbits have been scrambled by the merger. } 
\label{fig:kininsitunoagn}
\end{figure*}

\begin{figure*}
\centering
\includegraphics[width=\textwidth,height=\textheight,keepaspectratio]{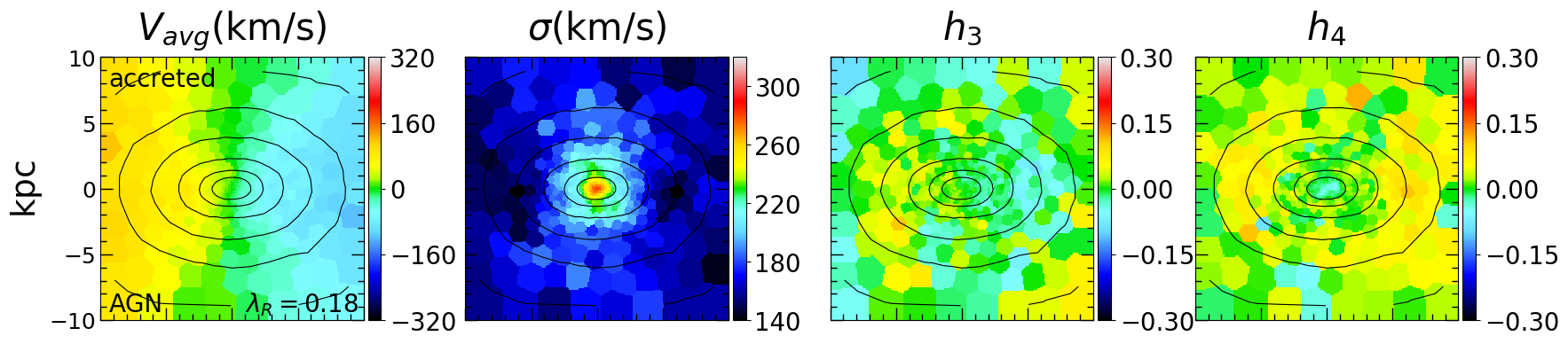} 
\includegraphics[width=\textwidth,height=\textheight,keepaspectratio]{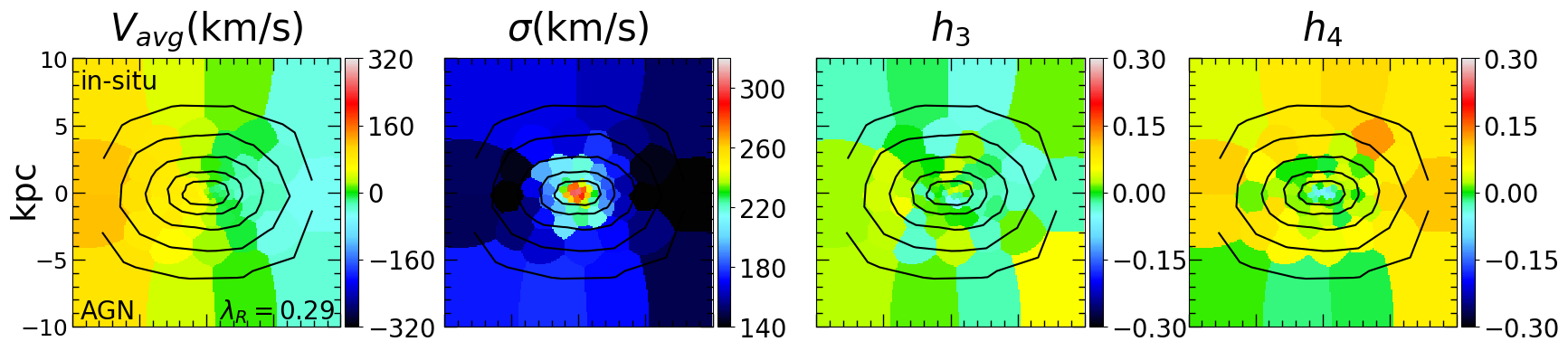}
\caption{Stellar kinematics of the accreted (above)
  and in-situ formed (below) stars of galaxy 0227 simulated with AGN
  feedback. The overall in-situ fraction is 17\% . In both cases the  
  kinematics are pressure-supported, as no star formation happened since 
  the last major merger at $z=0.25$. } 
\label{fig:kininsituagn}
\end{figure*}

\subsection{Kinematics of the accreted and in-situ-formed stellar components}
Our kinematic maps can be generated for different stellar components of the 
galaxy, to shed light on their 
respective kinematic structure. One might use the stellar age to
distinguish different components; we show this example in the appendix. 
Perhaps even more interesting though, is to separate stellar
particles according to their origin: either accreted from another
galaxy or formed in-situ in the main progenitor following the accretion 
of gas. Due to their 
intrinsically different origin, we can expect these two components 
to show very different kinematic (and stellar population) signatures 
(see e.g. \citealp{naab2014}). 
To classify stars as in-situ or accreted, we trace stars in the galaxies
throughout the simulation from $z=2$ to $z=0$, and
label them as in-situ stars when they form within ten per cent of 
the virial radius (see \citealp{oser2010}).
All the remaining stellar particles are labelled as accreted. 
In the case of galaxy 0227 the in-situ fraction is $f_{\rm in-situ} = 0.50$ 
and $f_{\rm in-situ} = 0.17$ for the \noagn\ and \agn\ cases, respectively. 
The values for the other galaxies are shown in Table \ref{tab:properties}. \newline
Figures \ref{fig:kininsitunoagn} and \ref{fig:kininsituagn} show the 
stellar kinematic maps obtained for the separated in-situ and accreted components, 
in the \noagn\ and \agn\ cases respectively. 
The accreted components (upper panels of Fig. \ref{fig:kininsitunoagn} and 
\ref{fig:kininsituagn}) exhibit a very high velocity dispersion in both
cases, but also have considerable net rotation, especially in the \noagn\ case.
This larger net rotation is probably caused by the potential being more
oblate-shaped in the \noagn\ simulation ($T=0.41$). In the \agn\ case the galaxy
has a very triaxial, almost prolate shape ($T=0.86$), which hinders the amount of z-tube
orbits (more on this in Section \ref{sec:orbit0227}) causing less rotation. \newline
The in-situ components are very different in the two cases. 
In the \agn\ case (lower panel of Fig. \ref{fig:kininsituagn}), the 
in-situ stars follow the same kinematics as the accreted ones. 
Almost all of these stars formed before the major merger at $z=0.25$, 
which means that their original orbits have been scrambled, resulting
in a dispersion-supported system.
In the \noagn\ case the number of in-situ-formed stars is larger, 
both before and after the major merger, and the corresponding kinematic 
maps are more complex. There are two distinct features. The first is an 
orderly fast-rotating disc in the midplane, with low velocity dispersion, a 
shallow \hthree\-\vavg/\vdisp trend, and strongly negative \hfour. The 
second is a slow-rotating bulge with high velocity dispersion and a much 
steeper trend with \hthree. The first component
is mostly made of young stars which formed after the $z=0.25$ major merger,
hence the orderly motion. The surrounding bulge is instead older. These 
stars formed in-situ at $z>0.25$, and their orbits have been scrambled
because of the major merger, resulting in less rotation. As the very high
velocity dispersion suggests, there is also a counter-rotating component
in this bulge, which explains why this component has a smaller net
rotation than the accreted stars in the same potential. \newline
This analysis implies that in-situ-formed stars and accreted stars 
tend to have intrinsically different kinematics from one another, at 
least until a major merger happens and scrambles their orbits. 
AGN feedback can thus significantly alter the present-day kinematics 
of galaxies by `freezing' the kinematics at the most recent major 
merger, affecting the orbits of both accreted and in-situ-formed stars.  

\subsection{Orbit distribution}\label{sec:orbit0227}
\begin{figure}
\centering
\includegraphics[width=\columnwidth,keepaspectratio]{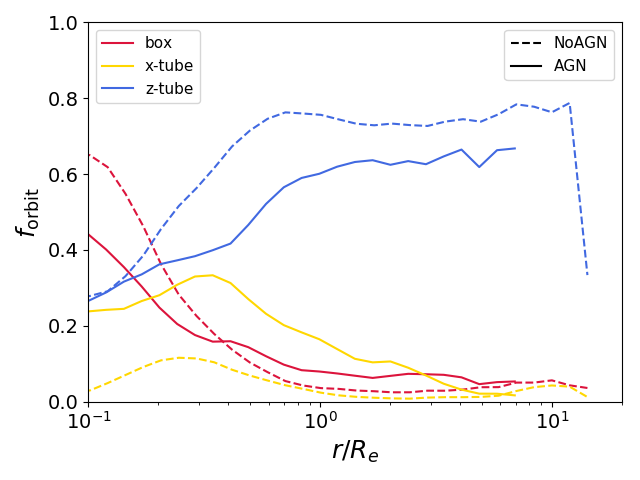} 
\caption{Radial frequency of three different types of orbits in our 
case study galaxy: z-tubes, x-tubes and boxes. The dashed line shows the 
\noagn\ case, and the full line shows the \agn\ one. In the latter, 
the fraction of z-tube orbits drops considerably due to the suppression 
of disc formation, and the fraction of x-tube orbits increases due to 
the more triaxial potential.} 
\label{fig:orbit0227}
\end{figure}
It is also of interest to directly study the distribution of 
stellar orbits, and how it is affected by AGN feedback.
We classify star particles into three global orbit types: z-tubes 
(rotating around the z-axis), x-tubes (rotating around the x-axis, including inner and outer major axis tubes)
and boxes (including $\pi$-boxes and boxlets). Figure \ref{fig:orbit0227} shows the fraction of 
these orbit families as a function of radius. In the 
\noagn\ case, the fraction z-tube orbits is larger at almost all radii. This  
is expected given the very prominent disc that has formed at low redshift. 
The central region is nevertheless dominated by box orbits, and x-tubes 
are very rare. In the \agn\ case there are significantly less z-tube 
orbits at all radii; the overall drop is from 65\% to 49\%, and the 
central regions are the ones that were impacted the most. The fraction 
of box orbits is slightly lower in the centre and higher in the outskirts. 
What really changed is the fraction of x-tube orbits, which went from 
an overall 5\% to 17\%. The likely reason for this is that the potential 
of the \agn\ galaxy has a more prolate shape ($T=0.86$, instead of $T=0.41$ 
for the \noagn\ case), allowing for this kind of orbits. This change in the balance of different orbit families also explains the positive correlation between \hthree\ and \vavg/\vdisp; the bulk of the LOS velocity distribution is made of x-tube, box and retrograde z-tube orbits, and the prograde z-tube orbits add a high-velocity tail to it. 

\section{Results from the simulation sample}\label{sec:gen}
So far we focused on a single, example galaxy. In this
section we show more general results for all twenty galaxies in our
sample. This analysis cannot reveal the statistical kinematic properties of 
quiescent galaxy populations from recent cosmological simulations
\citep{dubois2016,2017MNRAS.468.3883P,2017arXiv171201398L,2018arXiv180201583S}.
Instead, we would like to highlight the detailed impact of AGN feedback on massive
galaxies for a few individual systems simulated at higher resolution.
Table \ref{tab:properties} shows for each galaxy in our sample the stellar mass \Mstar\,
effective radius \Reff\, the average stellar age, the in-situ formed fraction, 
the ellipticity \ellip\, the isophotal shape \afour\ , the triaxiality parameter $T$, 
\lambdar\ , \hthreepar\ and the fraction of z-tube orbits $f_{\rm z-tube}$ . 
In general all our galaxies have a lower stellar mass with AGN feedback due to the quenching of 
star formation,  while the effective radius increases due to less dissipation (e.g. \citealp{2015MNRAS.450.1937C,2018arXiv180902143C}). In the following sections we will look at the distribution of kinematic (\lambdar\ , \hthreepar\ , orbit families) and morphological (\afour\ , triaxiality) properties at $z=1$ and $z=0$, and how AGN feedback affects them.
\newline 
\begin{table*}\label{tab:properties}
 \begin{center}
  \begin{tabular}{ | c | c | c | c | c | c | c | c | c | c | c | c |}
   \hline 
   GalID &    &$ M_\star (10^{10} M_\odot) $& $R_e$ & avg. age (Gyr) &  $f_{in-situ}$ & $\epsilon$ & \afour & $T$ & $\lambda_R$ & $ \hthreepar $ & $\fzpro\ $ \\ 
   \hline \hline

    0175 & NoAGN & $ 26.73 $ & $ 1.86 $ & $ 7.85 $ & $ 0.23 $ & $ 0.24 $ & $ 0.017 $ & $ 0.72 $ & $ 0.08 $ & $ -0.05 $ & $ 0.56 $\\ \hline 
    0175 & AGN & $ 18.93 $ & $ 2.57 $ & $ 10.72 $ & $ 0.10 $ & $ 0.35 $ & $ 0.001 $ & $ 0.51 $ & $ 0.12 $ & $ -6.58 $ & $ 0.45 $\\ \hline \hline
    0204 & NoAGN & $ 19.59 $ & $ 3.09 $ & $ 8.17 $ & $ 0.65 $ & $ 0.78 $ & $ 0.196 $ & $ 0.29 $ & $ 0.46 $ & $ -7.07 $ & $ 0.14 $\\ \hline 
    0204 & AGN & $ 16.41 $ & $ 2.06 $ & $ 9.50 $ & $ 0.31 $ & $ 0.37 $ & $ 0.026 $ & $ 0.15 $ & $ 0.36 $ & $ -8.19 $ & $ 0.38 $\\ \hline \hline
    0215 & NoAGN & $ 27.79 $ & $ 1.76 $ & $ 9.61 $ & $ 0.28 $ & $ 0.38 $ & $ 0.042 $ & $ 0.48 $ & $ 0.37 $ & $ -5.14 $ & $ 0.39 $\\ \hline 
    0215 & AGN & $ 7.38 $ & $ 1.70 $ & $ 11.26 $ & $ 0.12 $ & $ 0.31 $ & $ 0.018 $ & $ 0.50 $ & $ 0.02 $ & $ -0.79 $ & $ 0.15 $\\ \hline \hline
    0227 & NoAGN & $ 48.46 $ & $ 3.27 $ & $ 7.72 $ & $ 0.50 $ & $ 0.49 $ & $ 0.124 $ & $ 0.41 $ & $ 0.47 $ & $ -8.10 $ & $ 0.27 $\\ \hline 
    0227 & AGN & $ 22.24 $ & $ 2.60 $ & $ 9.95 $ & $ 0.17 $ & $ 0.15 $ & $ 0.000 $ & $ 0.86 $ & $ 0.10 $ & $ 0.50 $ & $ 0.62 $\\ \hline \hline
    0290 & NoAGN & $ 26.32 $ & $ 2.95 $ & $ 8.55 $ & $ 0.56 $ & $ 0.71 $ & $ 0.112 $ & $ 0.28 $ & $ 0.52 $ & $ -11.49 $ & $ 0.48 $\\ \hline 
    0290 & AGN & $ 12.67 $ & $ 2.57 $ & $ 10.45 $ & $ 0.29 $ & $ 0.38 $ & $ 0.045 $ & $ 0.66 $ & $ 0.06 $ & $ -2.62 $ & $ 0.38 $\\ \hline \hline
    0408 & NoAGN & $ 7.57 $ & $ 1.88 $ & $ 9.12 $ & $ 0.32 $ & $ 0.24 $ & $ 0.020 $ & $ 0.68 $ & $ 0.07 $ & $ 1.82 $ & $ 0.68 $\\ \hline 
    0408 & AGN & $ 15.98 $ & $ 2.59 $ & $ 8.84 $ & $ 0.58 $ & $ 0.43 $ & $ 0.055 $ & $ 0.32 $ & $ 0.36 $ & $ -6.41 $ & $ 0.27 $\\ \hline \hline
    0501 & NoAGN & $ 6.80 $ & $ 1.74 $ & $ 10.71 $ & $ 0.14 $ & $ 0.33 $ & $ 0.029 $ & $ 0.22 $ & $ 0.42 $ & $ -6.74 $ & $ 0.59 $\\ \hline 
    0501 & AGN & $ 8.25 $ & $ 1.93 $ & $ 11.22 $ & $ 0.16 $ & $ 0.30 $ & $ 0.018 $ & $ 0.44 $ & $ 0.10 $ & $ -3.55 $ & $ 0.62 $\\ \hline \hline
    0616 & NoAGN & $ 8.61 $ & $ 1.29 $ & $ 8.87 $ & $ 0.30 $ & $ 0.38 $ & $ 0.055 $ & $ 0.09 $ & $ 0.04 $ & $ -1.02 $ & $ 0.17 $\\ \hline 
    0616 & AGN & $ 4.56 $ & $ 1.53 $ & $ 11.07 $ & $ 0.09 $ & $ 0.35 $ & $ 0.010 $ & $ 0.36 $ & $ 0.35 $ & $ -7.27 $ & $ 0.08 $\\ \hline \hline
    0664 & NoAGN & $ 8.04 $ & $ 1.15 $ & $ 9.50 $ & $ 0.35 $ & $ 0.41 $ & $ 0.031 $ & $ 0.37 $ & $ 0.44 $ & $ -5.74 $ & $ 0.69 $\\ \hline 
    0664 & AGN & $ 7.23 $ & $ 1.38 $ & $ 10.51 $ & $ 0.22 $ & $ 0.43 $ & $ 0.052 $ & $ 0.09 $ & $ 0.32 $ & $ -7.61 $ & $ 0.56 $\\ \hline \hline
    0858 & NoAGN & $ 3.67 $ & $ 2.25 $ & $ 8.97 $ & $ -1.00 $ & $ 0.36 $ & $ 0.045 $ & $ 0.25 $ & $ 0.49 $ & $ -6.02 $ & $ 0.49 $\\ \hline 
    0858 & AGN & $ 6.99 $ & $ 1.93 $ & $ 8.11 $ & $ 0.49 $ & $ 0.26 $ & $ 0.006 $ & $ 0.38 $ & $ 0.31 $ & $ -7.20 $ & $ 0.37 $\\ \hline 
  \end{tabular}
 \caption{General properties of our sample of simulated galaxies. }
 \end{center}
\end{table*}

\subsection{Angular momentum}
\begin{figure*}
\centering
\includegraphics[width=\columnwidth]{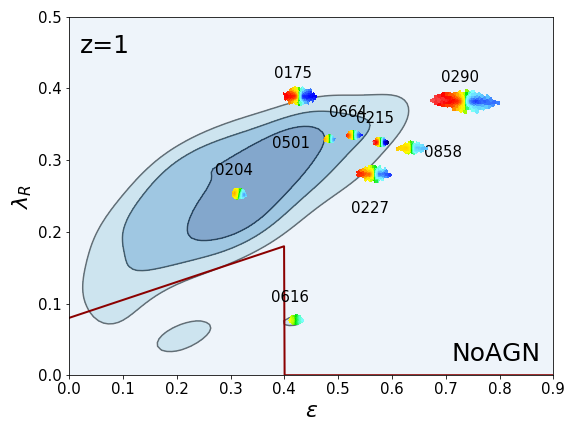} 
\includegraphics[width=\columnwidth]{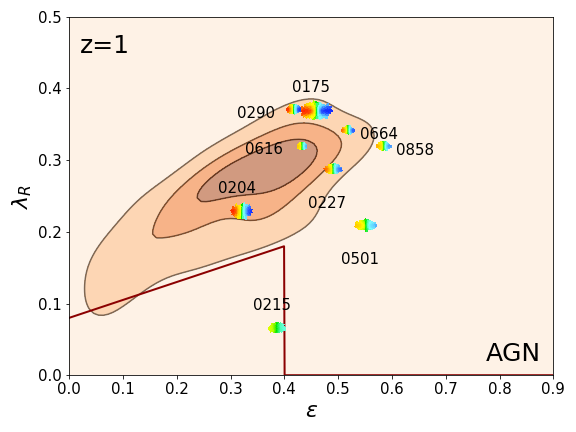}
\includegraphics[width=\columnwidth]{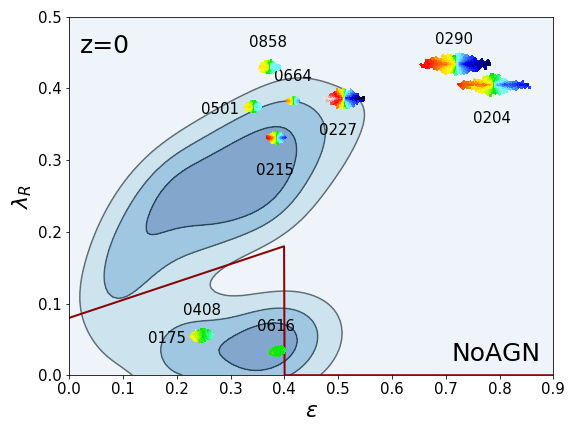}
\includegraphics[width=\columnwidth]{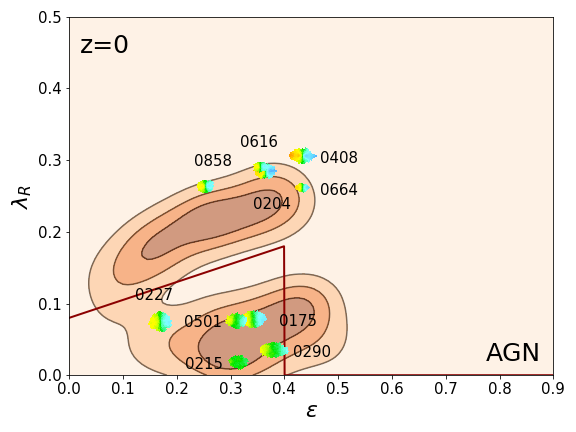}
\caption{\lambdar\ as a function of ellipticity \ellip\ for the
  galaxies at redshift $z=1$ (top panels) and $z=0$ (bottom panels),
  simulated without (\noagn, left) and with (\agn, right) AGN 
  feedback. The edge-on values are indicated by velocity maps. The
  coloured contours indicate the distribution of our galaxies when
  they are seen from 50 random orientations each. The dark red line 
  marks the limit between slow- and 
  fast-rotators according to \citet{cappellari2016}. With AGN feedback
  the systems become rounder and rotate more slowly at z=0. }
\label{fig:lambdaell} 
\end{figure*}
\begin{figure}
\centering
\includegraphics[width=\columnwidth]{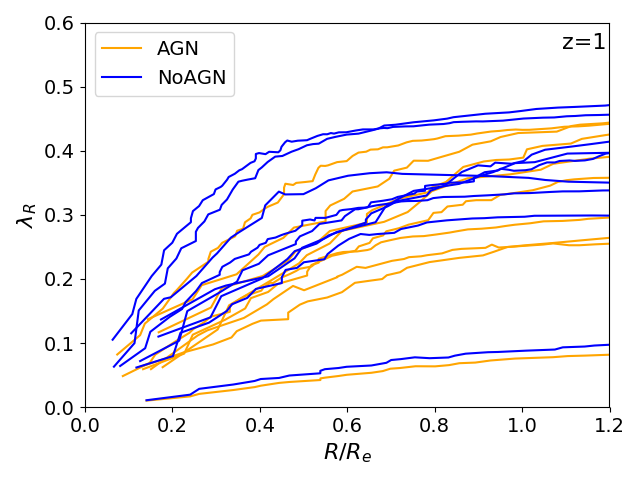}
\includegraphics[width=\columnwidth]{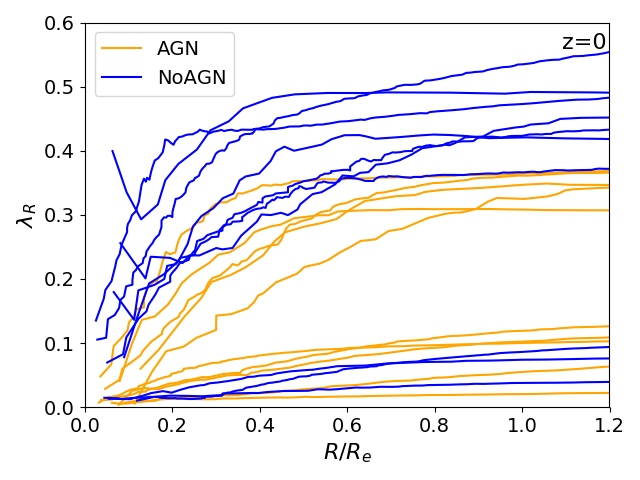}
\caption{$\lambda_R$ radial profiles of our galaxy sample at z=1 (top)
  and z=0 (bottom). The \agn\ galaxies (orange) evolve towards lower 
  $\lambda_R$ values than their \noagn\ counterparts (blue).} 
\label{fig:lambdarprof}
\end{figure}
 
In Fig. \ref{fig:lambdaell} we plot the \lambdar\ parameter of the sample
galaxies versus their ellipticity \ellip\ for the simulations without 
(\noagn, left panels) and with AGN
(\agn, right panels) at redshift $z=1$ (top panels) and $z=0$ (bottom
panels). The location of edge-on projections are indicated by the
velocity maps. The blue/orange shaded regions indicate the typical 
distribution of these systems for random orientations (projection effects for 
\lambdar\ based on simulations are discussed in
e.g. \citealp{jesseit2009,naab2014,lagos2018}). 
They were obtained by calculating \lambdar\ and \ellip\ for 50 random lines-of-sight for each galaxy.
The red line separates slow- and fast-rotators following
to the definition by \citet{cappellari2016}. 
A galaxy is considered a slow-rotator when 
\begin{equation} \label{lambdacrit}
\lambdar < 0.08 + 0.25 \, \ellip {\hspace{5mm} \rm  with \hspace{5mm}} \ellip < 0.4 .
\end{equation}
The distribution of galaxies at $z=1$ is similar between the \agn\ and 
\noagn\ cases, with most galaxies being flattened fast-rotators 
with \lambdar\ in the range 
$0.2 < \lambdar < 0.4$. The ellipticity values are a bit higher in the 
\noagn\ case ($0.3 < \ellip < 0.8$) than in the \agn\ one ($0.3 < \ellip < 0.6$), 
but qualitatively the two populations are very similar. 
By $z=0$ many (7 out of 10) of the \noagn\ galaxies are still fast
rotators with a similar ellipticity distribution. 
This trend is in agreement with results for massive galaxy populations from
cosmological box simulations without AGN feedback \citet{dubois2016}.
Instead, in the \agn\ case by $z=0$ the galaxies have 
become rounder ($\epsilon < 0.4$) and more slowly
rotating, with \lambdar\ no larger than $\sim 0.35$. More than half of
the galaxies would be considered bona-fide slow rotators even in their
edge-on projections.
As discussed earlier, the trend towards slower
rotation with AGN feedback is caused by the suppression of late in-situ 
star formation (see \citealp{2018ApJ...860...14B} for a discussion of 
ejective and preventative AGN feedback), 
which in most cases significantly reduces rotation
observed at $z=0$. The effect is strongest for the largest and most
massive galaxies in our samples (lower numbers, like 0227), which - 
without AGN feedback - develop massive fast-rotating disc structures
 (In the case of galaxy 0175, this young 
disc structure is on a different plane, and thus does not increase 
\lambdar\ significantly).

We find a correlation between \lambdar\ and \ellip, at least for the
fast-rotators: faster rotating galaxies tend to be more
flattened. Most of our slow-rotating galaxies exhibit a relatively
high ellipticity, which is a trend found in other simulation studies
as well\citep{bois2010,naab2014}, and possibly due to resolution limits.
An interesting case is galaxy 0616, which contradicts our expectations
by being a slow-rotator when simulated without AGN feedback but turns
into a fast-rotator when simulated with AGN feedback. What happens
here? In the \noagn\ case gas infall triggers a starburst that forms 
a disc that counter-rotates with respect 
to the rest of the galaxy. This lowers the projected
\lambdar\ value, but leaves a relatively high ellipticity. 
In the \agn\ case the gas is kept from forming this new disc and 
the galaxy retains most  of the (projected) angular
momentum of the older stellar component. \newline
In Figure  \ref{fig:lambdarprof} we plot the \lambdar\ radial profiles 
for all galaxies, at $z=1$ and $z=0$. Typically, the values increase 
from the centre until they reach an asymptotic value, usually within
$R_e$. This is consistent with previously published simulation data,
even though we are missing systems with dropping \lambdar\ profiles 
\citep{naab2014,2014MNRAS.438.2701W,lagos2018}. 
At $z=1$ there is not much difference 
between the \agn\ and \noagn\ galaxies, while at $z=0$ 
galaxies simulated with AGN feedback show once again systematically
lower \lambdar\ values, even among the fast-rotators. 
Many galaxies that would be
rotationally-supported without AGN, become pressure-supported when an AGN is
present.  Overall, AGN feedback results in more slow-rotating and dispersion-supported galaxies 
in agreement with previous simulations \citep{dubois2016} and the statistics
of observed early-type galaxies. 

\subsection{Higher-order kinematics and orbital structure} \label{sec:hthreepar}
\begin{figure*}
\centering 
\includegraphics[width=\columnwidth]{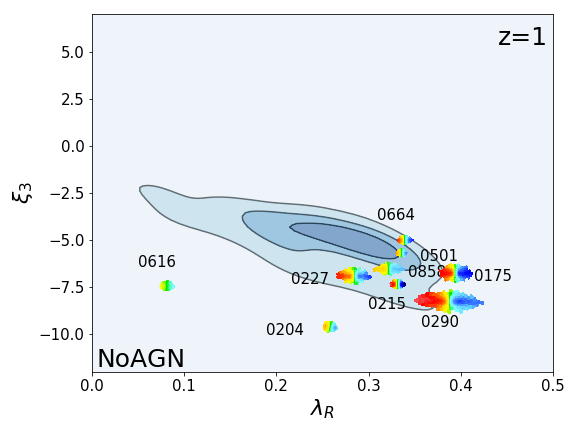} 
\includegraphics[width=\columnwidth]{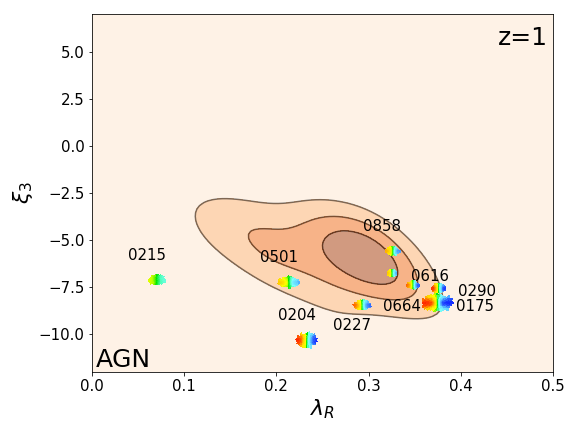}
\includegraphics[width=\columnwidth]{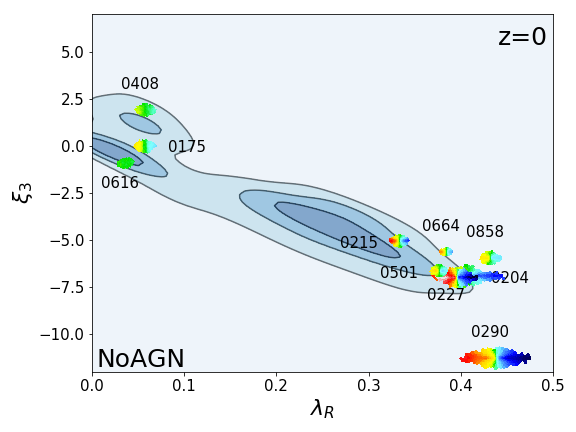}
\includegraphics[width=\columnwidth]{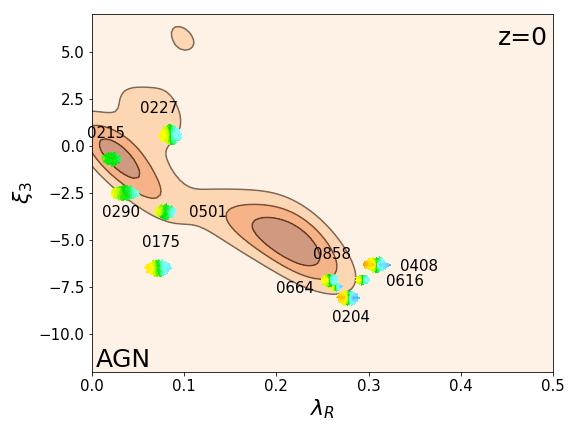}
\caption{\hthreepar\ versus \lambdar\ at $z=1$ (top
  panels) and $z=0$ (bottom panels), simulated without (left) and with
  (right) AGN feedback. The kinematic map markers indicate the values 
when the galaxy is seen edge-on, while the density contours indicate 
the distribution when our galaxies are seen through 50 random orientations each. 
At $z=1$ all galaxies have have values of \hthreepar\ in the anti-correlation  
regime, typical of fast-rotators, while at $z=0$ many galaxies have  
$\hthreepar\ \sim 0$ or in a few cases even positive, and this effect is 
stronger with AGN feedback.} 
\label{fig:lambdah3}
\end{figure*}

\begin{figure}
\centering
\includegraphics[width=\columnwidth]{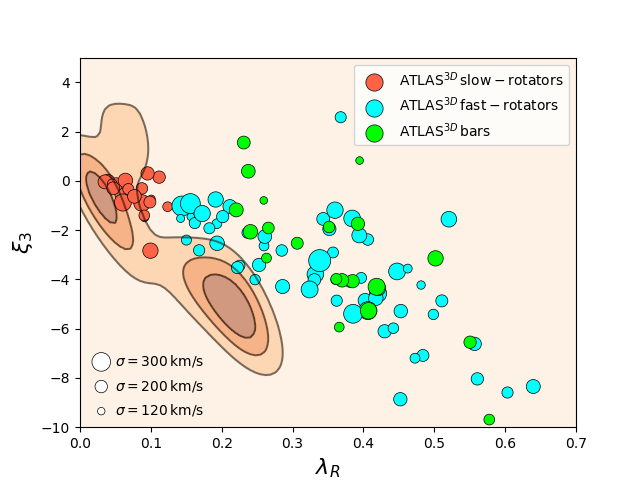}
\caption{\hthreepar\ as a function of \lambdar\ for the galaxies from
  the \atlas\ sample (circle markers), compared with our \agn\ simulations 
  (orange contours; same as Figure \ref{fig:lambdah3}). The \atlas\ galaxies
  are distinguished in slow-rotators (red) and fast-rotators (light blue) 
  according to the \citet{cappellari2016} definition. At equal \lambdar, observed fast-rotators
  seem to have smaller \hthreepar\ (absolute) values than the simulation, possibly because of more
  complex kinematic features (bars) and of more noisy \hthree\ measurements. Slow-rotators have $\lambdar \sim 0$ and $\hthreepar \sim 0$ in both observations and simulations.} 
\label{fig:atlas}
\end{figure}

As discussed in sections \ref{sec:h3def} and \ref{sec:angiso}, 
rotating galaxies are expected to have anti-correlated \hthree\ and 
velocity fields, but the degree of this anti-correlation depends on the 
orbital structure of the galaxy, and we can employ our \hthreepar\ 
parameter to evaluate this for our sample. 
In Figure \ref{fig:lambdah3} we plot \hthreepar\ as a function of 
\lambdar\ at $z=1$ and $z=0$. The edge-on values
are plotted with velocity maps, while the contours represent the
location of the sample in the \hthreepar - \lambdar\ plane for random orientations. 
Generally, the edge-on \hthreepar\ values are larger in absolute value, but for 
different inclinations the dependence of \hthreepar\ on the viewing angle is weak.
At $z=1$ all galaxies have a negative of \hthreepar\ and \hthree\ is anti-correlated 
with the velocity, as expected for fast-rotators. This is also 
true for the two galaxies which are slow-rotators (according to 
\lambdar) at $z=1$. At $z=0$ the sample 
splits into two groups: slow-rotators with low values of \lambdar\ tend 
to have $\hthreepar \sim 0$ (very steep correlation or no correlation), 
while all fast-rotators have $\hthreepar\ < -3$ (negative correlation). 
The specific value of \hthreepar\ for the fast-rotators depend on their 
orbital structure; the galaxies where a disc feature is particularly 
prominent (0204, 0227 and 0290 in the \noagn\ case) have the lowest values, 
reaching about $\hthreepar = -11.5$. In other words, more flattened and simple 
rotating systems 
have a less steep correlation between \hthree\ and \vavg/\vdisp\ than 
fast-rotators with more complex kinematics. A similar behaviour was also 
observed in real galaxies by \citet{veale2017}. This results in a weak 
correlation between \hthreepar\ and \lambdar\ for the fast-rotators, that 
was not present at $z=1$ when the kinematics of the galaxies were overall simpler.
The bi-modality of slow- and fast-rotators in the \hthreepar-\lambdar\ plane 
is seen in both the \noagn\ and \agn\ cases, but with AGN feedback the group 
of galaxies with $\hthreepar \sim 0$ is larger. A few galaxies have a 
positive value of \hthreepar\ at $z=0$. One of them, 0227 \agn, has already 
been extensively discussed. The other one, 0408 \noagn, has a positive value 
because of a sub-dominant rotating component in an otherwise 
dispersion-supported system, producing a positive correlation between \hthree\ 
and \vavg.   \newline
If we compare these results with observational IFU surveys, we find a small
discrepancy. In Figure \ref{fig:atlas} we plot the \hthreepar\ values of
galaxies from the \atlas\ survey \citep{atlasoverview}\footnote{Available from 
http://purl.org/atlas3d},
compared with the contours of our \agn\ simulations seen at random inclinations. 
The \atlas\ values also 
include a re-extraction of the kinematics from the subset of galaxies in the 
SAURON survey originally presented in \citet{emsellem2004}.
To compute \hthreepar\ and \lambdar\ for the \atlas\ sample, 
we only considered spaxels with $\sigma > 120 \rm km/s$, since the Gauss-Hermite moments can only be extracted 
from the data when the galaxy velocity dispersion is well resolved by 
the spectrograph (e.g. \citealp{2004PASP..116..138C}). 
The distribution of \hthreepar\ values is similar between observations
and simulations, and can be divided in two groups: slow-rotators with 
$\hthreepar \sim 0$ and fast-rotators with $-3 < \hthreepar <-10$. 
However, at given \lambdar the \atlas\ galaxies seem to have lower \hthreepar 
(in absolute value) than the simulations. We believe there are at least 
three reasons for this difference.
Several of the \atlas\ fast-rotators have strong bar features, which
are not present in our sample of simulations. In their presence the kinematic
maps often show a positive correlation between \vavg\ and \hthree\ 
\citep{2004AJ....127.3192C}, causing 
\hthreepar\ values closer to zero or sometimes even positive. 
In Figure \ref{fig:atlas} galaxies with clear bars have been highlighted, 
but hidden or weak bars could be present in the other galaxies too, affecting 
the \hthree\ values. Secondly, 
as previously mentioned, constraining the \hthree\ value of each spaxel is harder in observations. 
The selection of spaxels with $\sigma > 120 \rm km/s$ limits this problem, but does not eliminate it.
This results in more noisy \hthree\ maps, which makes the \hthree-\vavg/\vdisp\ 
trend less tight, and thus moves the \hthreepar\ value of observed galaxies 
closer to zero. At equal $\sigma$, this effect is stronger for slower-rotating galaxies, 
as their LOS velocity distribution have lower \hthree\ values.
Lastly, our (\agn) sample consists of only 10 massive galaxies, all of which have 
relatively low \lambdar values. This means that our simulations do not explore 
the $\lambdar > 0.3$ regime, but if they did, we would expect most of them 
to have $-10 < \hthreepar < -5$, like many of the galaxies in our \noagn\ sample, 
matching the observations. 

\subsection{Orbit distribution and \hthreepar}\label{sec:orbitsample}
\begin{figure}
\centering
\includegraphics[width=\columnwidth]{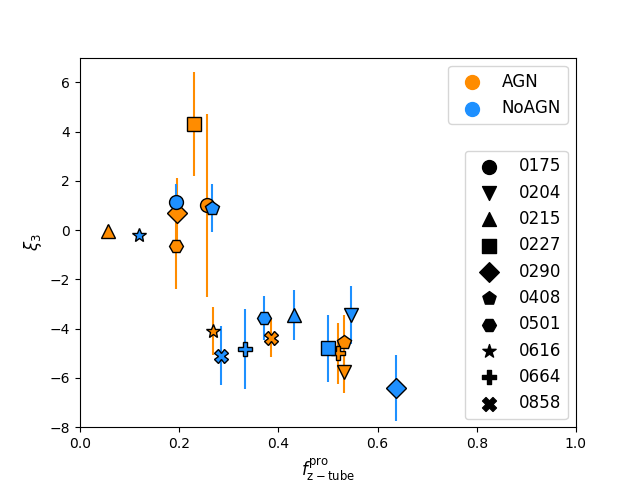}
\caption{\hthreepar\ as a function of the fraction of prograde z-tube 
orbits \fzpro\ for our sample of simulated galaxies at $z=0$. 
The \hthreepar\ values of each galaxy are an average over 50 random views, 
and the error bars are their standard deviation.
Galaxies with high \fzpro\ tend to have $\hthreepar < -5$. } 
\label{fig:orbith3}
\end{figure}

\begin{figure}
\centering
\includegraphics[width=\columnwidth]{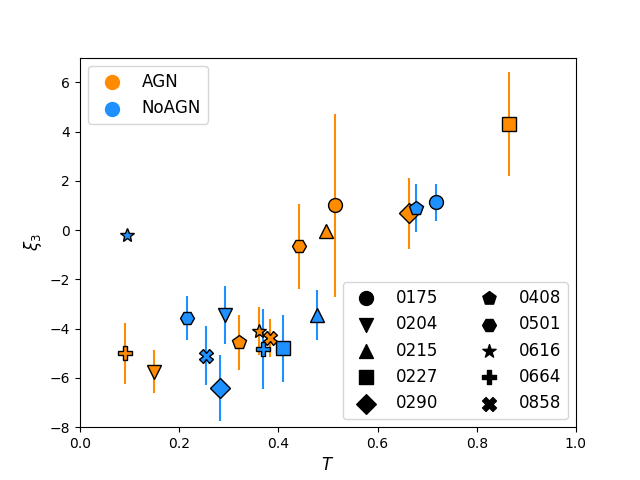}
\caption{\hthreepar\ as a function of the triaxiality parameter for our 
sample of simulated galaxies at $z=0$. 
The \hthreepar\ values of each galaxy are an average over 50 random views, 
and the error bars are their standard deviation.
There is a weak correlation between the two parameters: 
prolate galaxies have positive values of \hthreepar\ , while oblate
galaxies have negative values. } 
\label{fig:h3triax}
\end{figure}

\begin{figure}
\centering
\includegraphics[width=\columnwidth]{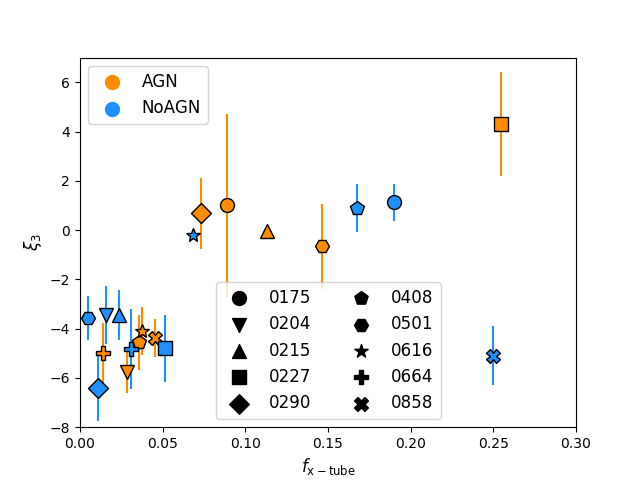}
\caption{\hthreepar\ as a function of the fraction of x-tube 
orbits \fxtube\ for our sample of simulated galaxies at $z=0$.
The \hthreepar\ values of each galaxy are an average over 50 random views, 
and the error bars are their standard deviation.
Galaxies with high \fxtube\ tend to have $\hthreepar > 0$. } 
\label{fig:orbith3x}
\end{figure}

We would also like to see how closely connected \hthreepar\ is to the actual
orbital structure of galaxies, measured in the same way as in Sections \ref{sec:orbitan} 
and \ref{sec:orbit0227}. In Figure \ref{fig:orbith3} we plot \hthreepar\ as a 
function of the fraction of prograde z-tube orbits within \Reff, \fzpro, 
at $z=0$. 
The plotted \hthreepar\ values are the average for 50 random views of each galaxy, 
and the error bars mark the dispersion (negligible for galaxies 0616 \noagn\ and 0215 \agn). 
Most galaxies with high values of \fzpro\ have a $\hthreepar < -3$ 
as expected, and there is 
a rough correlation between the two quantities. The galaxy with the highest 
\fzpro\ (0290 \noagn) is also the one with the lowest value of \hthreepar : 
$\sim -11.5$ when seen edge-on and $\sim -6.5$ when averaging between many different viewing angles. The reason for this is that when the system is 
dominated by orbits that rotate (progradely) around the z axis, these 
stars form the bulk of the LOS velocity distribution, and all other 
orbit types make the \hthree\ signal stronger for that given \vavg/\vdisp. 
When non-rotational orbits are dominating ($\fzpro\ \sim 0$), then $ 
\vavg/\vdisp \sim 0$ and consequently $\hthreepar\ \sim 0$. \newline
A few galaxies (0175 \noagn, 0408 \noagn\ and 0227 \agn ) have a positive
correlation between \hthree\ and \vavg/\vdisp\ in large parts of their 
kinematic maps, resulting in a positive value of \hthreepar. This is 
likely connected to the fact that these galaxies have a prolate potential.
We investigate this by plotting \hthreepar\ as a function of the triaxiality
parameter \triax\ in Figure \ref{fig:h3triax}. There seems to be a rough
correlation between the two quantities in our sample. The most prolate galaxies
($\triax \sim 1$) have positive values of \hthreepar, while almost all oblate 
galaxies ($\triax << 1$) have negative values. The one exception is galaxy
0616 \noagn, which as already discussed is made of two counter-rotating 
components and looks like a `fake' slow-rotator. \newline
This connection between morphology and kinematics likely arises because
different potential shapes allow different kinds of orbits; specifically,
x-tubes are more common in prolate potentials. We see this by plotting
\hthreepar\ as a function of the fraction of x-tube orbits \fxtube\ in 
Figure \ref{fig:orbith3x}. There is again a rough correlation, meaning that
galaxies with higher \fxtube\ are more likely to display a positive
correlation between \hthree\ and \vavg/\vdisp\ in their kinematic maps.
This follows from the correlation between \fxtube\ and the triaxiality \triax, 
which has previously been observed in isolated \citep{2005MNRAS.360.1185J} 
and cosmological simulations \citep{roettgers2014}. It should however be noted that in a pure prolate system only x-tube orbits and box orbits are allowed, and if there is net rotation around the long axis  \hthree\ and \vavg/\vdisp\ become anti-correlated again. We do not see this in our sample because none of our galaxies is dominated by x-tube orbits (at most $\fxtube = 0.25$, for 0227 \agn ).

\subsection{Isophotal shape}
\begin{figure*}
\centering
\includegraphics[width=\columnwidth]{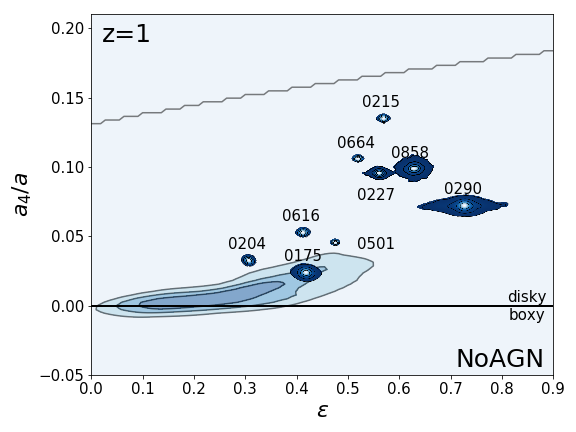} 
\includegraphics[width=\columnwidth]{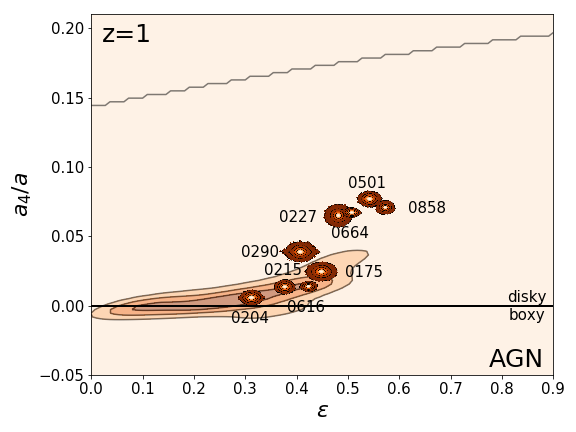}
\includegraphics[width=\columnwidth]{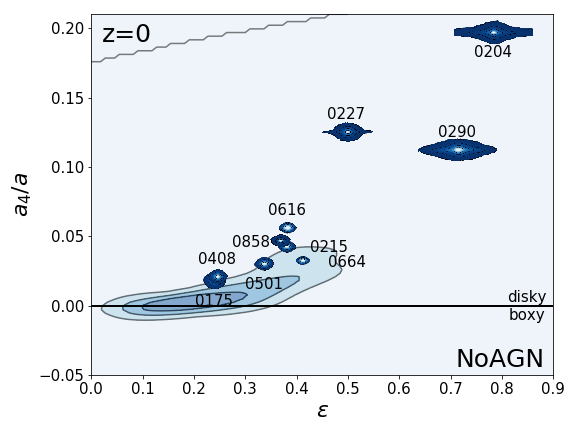}
\includegraphics[width=\columnwidth]{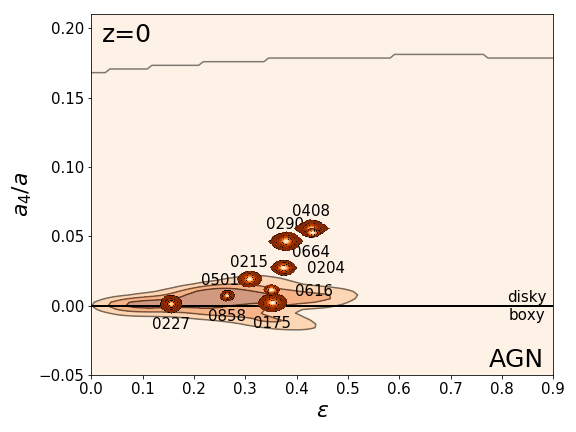}
\caption{\afour\ versus galaxy ellipticity \ellip\ at $z=1$ (top
  panels) and $z=0$ (bottom panels), simulated without (left) and with
  (right) AGN feedback. The edge-on locations are indicated by the
  isophotal maps, while the density contours indicate the distribution 
  of our galaxies when they are seen from 50 random orientations each. 
  The black line indicates elliptical isophotes and
  separates boxy ($a_4 < 0$) from disky ($a_4 > 0$) galaxies. Galaxies
  with AGN feebdack are rounder and have more elliptical - in one case
  even boxy - isophotes at $z=0$. } 
\label{fig:a4aell}
\end{figure*}

In Fig. \ref{fig:a4aell} we plot the \afour\ parameter of all our
galaxies versus their ellipticity \ellip at $z=1$ and $z=0$. Like for
Figs. \ref{fig:lambdaell} and \ref{fig:lambdah3}, we also added contours 
to show the distribution of values for smaller inclinations. At
$z=1$ the panels with and without AGN feedback look qualitatively very
similar. All galaxies have disky isophotes when viewed edge-on. When
viewing the galaxies from different points of view both the
ellipticity and the \afour\ values tend to become smaller. 
At $z=0$, the cases with and without AGN behave as expected. 
The \noagn\ galaxies show systematically higher \afour\ values, 
due to the formation of embedded stellar discs at low redshift. 
In the \agn\ case the \afour\ values are lower, meaning that the 
isophotes are less disky and closer to elliptical. Even
though we do not have a clearly boxy galaxy in our sample, two
galaxies (0175 and 0227) have almost perfectly elliptical isophotes.  \newline
We also computed the
three-dimensional shape of our galaxies using the triaxiality parameter $T$, defined
in Section \ref{sec:photo}. 
The values of $T$ for our galaxies are found in table \ref{tab:properties}, or in Figure \ref{fig:h3triax}. 
We found that with AGN feedback a
bigger fraction of our galaxies (five out of ten, instead of two out of ten) has a triaxial or almost prolate shape ($T > 0.5$). A prolate shape
is more common for massive ellipticals, as found in both observations
\citep{tsatsi2017,krajnovic2018,2018MNRAS.477.4711G} and simulations \citep{lidandan2017}. 
Without AGN feedback more of our galaxies are oblate ($T \sim 0$) 
despite their larger mass, which makes them more similar to the significantly 
less massive fast-rotators we observe \citep{krajnovic2011,cappellari2016}. 

\section[]{Discussion and conclusions}\label{sec:conc}
From the analysis of these simulated galaxies emerges a clear picture,
which confirms the previous studies on the subject and adds new insights. 
The energy output of AGNs heats up and pushes away the
interstellar gas, effectively suppressing the in-situ formation of
stars. This affects the kinematics and morphology of the systems with 
a stronger impact at later cosmic times, when the central black 
holes become more massive. In our simulations AGN feedback results 
in realistic early-type galaxy properties at $z=0$. 
From our detailed stellar assembly, stellar population, 
mock IFU, isophotal shape and stellar orbit analysis we get the following 
generic picture: 
\begin{itemize}
\item The stellar kinematics of massive early-type galaxies is 
  significantly affected by AGN
  feedback, as seen both in the mock observational kinematic maps and 
  in the orbit analysis of our simulation. Without
  AGN feedback massive early-type galaxies would develop young
  fast-rotating stellar discs even at low redshift, giving them
  kinematic signatures typical of less massive fast-rotators. With AGN
  feedback massive early-type galaxies are instead more likely to become 
  slow-rotators due to the suppression of late in-situ star formation,
  in agreement with previous studies 
  \citep{2013MNRAS.433.3297D, 2014MNRAS.443.1500M, 2017MNRAS.468.3883P,lagos2018}.

\item As shown in Figure \ref{fig:lambdakinz}, the slowing-down effect 
  of AGN feedback is more pronounced in, 
  but not limited to, late major mergers. Apart for some cases where
  mergers can cause a spin-up of the galaxy thanks to a favourable orbital
  configuration \citep{naab2014}, most of the time mergers tend to
  disrupt the orbits of stars, reducing the angular momentum of the galaxy. 
  However, without AGN feedback the further accretion of gas can produce a
  new rotating stellar disc and make the galaxy recover its angular
  momentum. With AGN feedback the in-falling star-forming gas is
  heated up and blown away. 
  The origins of this mechanism lie in the different
  spatial and kinematic properties of in-situ-formed and accreted stars 
  \citep{2016MNRAS.458.2371R}.
 
\item AGN feedback starts having a significant impact on the stellar 
  angular momentum only after $z=1$, and is stronger for more massive galaxies. 
  With some exceptions, like galaxy 0616 in our
  sample which without AGN feedback develops a counter-rotating core,
  having AGN feedback always decreases the angular momentum of the
  galaxies in our sample.  

\item We compute the ellipticity \ellip\ and the \afour\ isophotal
  shape parameter and follow their evolution through cosmic time. 
  By suppressing the formation of discs, AGN feedback 
  makes galaxies less flattened and their isophotes significantly
  less disky (more elliptical or even boxy), especially when seen edge-on. 
  Like for the angular
  momentum, this difference starts arising at $z \sim 1$, and its 
  effect is again stronger for the most massive galaxies of our 
  sample. 

\item We introduce a new global parameter, \hthreepar, to quantify 
  the anti-correlation between the LOS-velocity and 
  \hthree\ from two-dimensional kinematic maps. Slow- and fast-rotators have
  different typical values of this parameter, owing to their different
  orbital structures. AGN feedback pushes the
  \hthreepar\ value towards the slow-rotator
  regime ($\hthreepar \sim 0$, meaning a very steep anti-correlation 
  between \vavg/\vdisp\ and \hthree\ or lack of such a correlation). 

\item We perform a full orbit analysis for all simulated galaxies 
  and find that systems with AGN feedback have a higher fraction 
  of x-tube and box orbits and a lower fraction of z-tubes. 
  This is consistent with them being more triaxial due to the lack 
  of late in-situ star formation and the more stellar accretion 
  dominated assembly history. 
  We find that the \hthreepar\ parameter is well correlated to the
  fractions of prolate z-tubes and x-tubes, as well as with the
  triaxiality of the galaxy.

\item We compared the \hthreepar\ values of our simulations with observed
  galaxies from the \atlas\ sample, finding an interesting discrepancy. 
  At equal \lambdar, observed fast-rotators seem to have values of 
  \hthreepar\ closer to zero and sometimes even positive; this could 
  be because many of these galaxies show bar features, which cause 
  a positive correlation between \hthree\ and LOS velocity, and/or 
  possibly because of noise in the observed \hthree\ values. Our 
  \agn\ sample also lacks galaxies with high \lambdar\ values, which
  are instead very common in the \atlas\ sample.  
  
\end{itemize}
Even though slow-rotating galaxies could also form
without AGN feedback through particularly gas-poor formation paths,
our simulations suggest that AGN feedback might
be essential to produce the observed amount of quiescent,
slow-rotating and non-disky early-type galaxies. 
The impact of AGN on the rotation properties are in line with earlier studies using different AGN feedback models and 
simulation codes \citep{2013MNRAS.433.3297D, 2014MNRAS.443.1500M, 2017MNRAS.468.3883P,lagos2018}. In this study we indicate that also higher-order properties in the isophotal shape and line-of-sight 
kinematics, as well as the underlying orbital content, are significantly affected by accreting supermassive black holes. The effects typically results in a better 
agreement with observations. The newly introduced kinematic asymmetry parameter \hthreepar\ might provide a useful diagnostic for large integral field surveys, as it is a kinematic indicator for intrinsic shape and orbital content.
This study 
is not statistically complete nor can the assumed AGN feedback model 
be considered as an accurate description of the process. We can just 
give a model perspective on the observable effect of processes 
eventually happening in nature. 

\section*{Acknowledgments}
This research was supported by the German Federal Ministry of Education and Research (BMBF) within the German-South-African collaboration project 01DG15006 "Ein kosmologische Modell f\"ur die Entwicklung der Gasverteilung in Galaxien". 
TN acknowledges support from the DFG Excellence Cluster "Universe". 
MH acknowledges financial support from the European Research Council (ERC) via an Advanced Grant under grant agreement no. 321323-NEOGAL.

\bibliographystyle{mnras}
\bibliography{mf}

\appendix
\section[]{Additional figures}
\begin{figure*}
\centering  
\includegraphics[width=\textwidth]{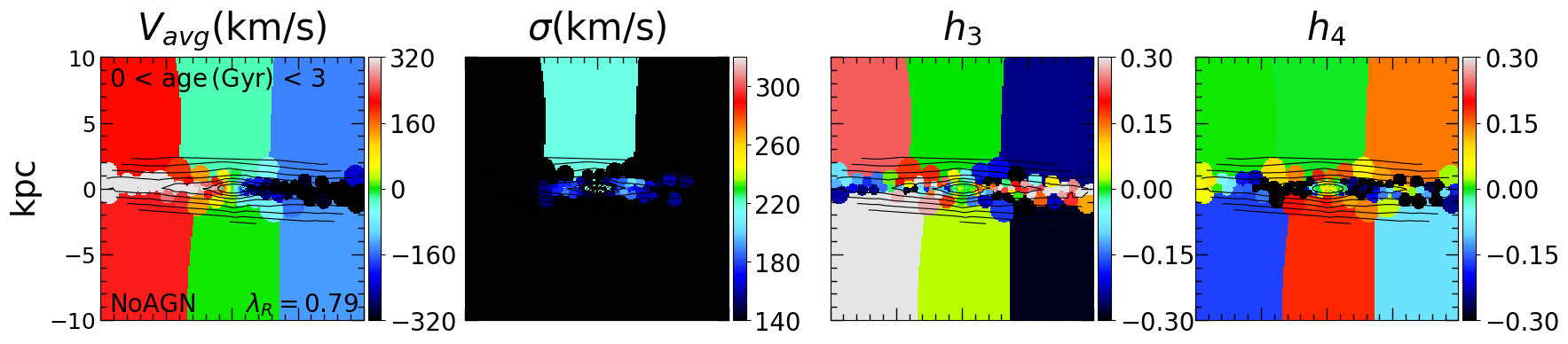} 
\includegraphics[width=\textwidth]{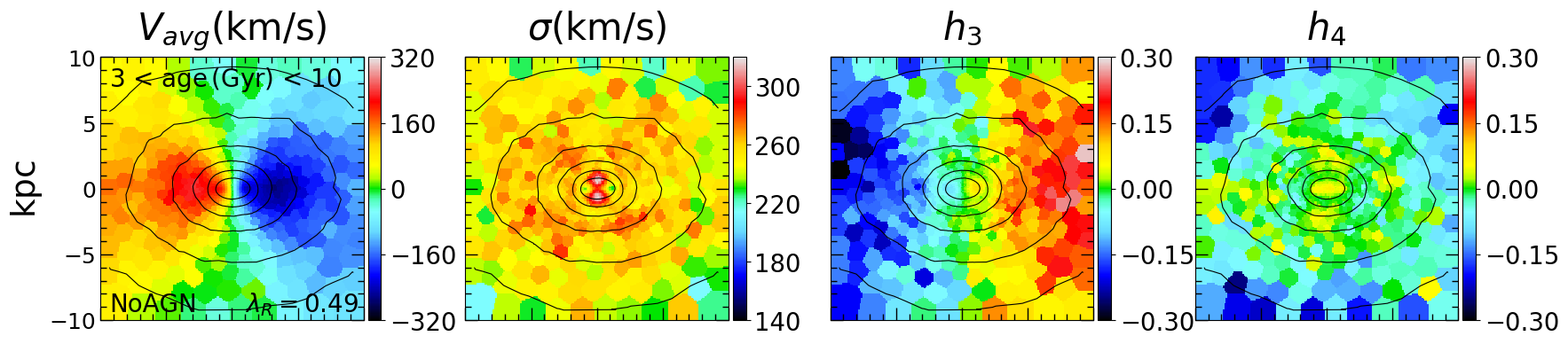}
\includegraphics[width=\textwidth]{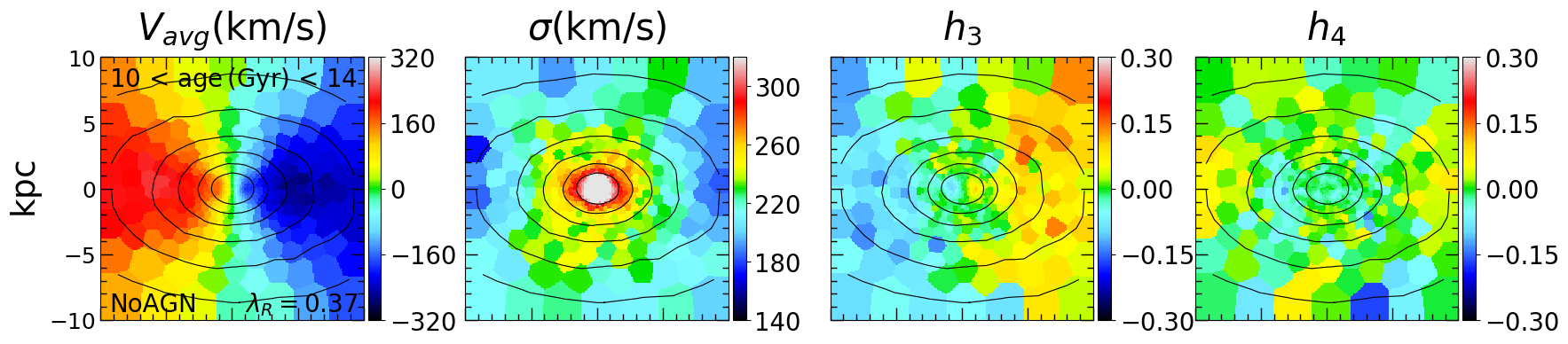}

\caption{Kinematic maps (\vavg,\vdisp,\hthree,\hfour) for age-selected 
stellar components of galaxy 0227 simulated without AGN. From top to bottom, 
stars between 0 and 3 Gyr old, between 3 and 10 Gyr old, and older than 10 Gyr. }
\label{fig:kinagenoagn}
\end{figure*}

\begin{figure*}
\centering
\includegraphics[width=\textwidth]{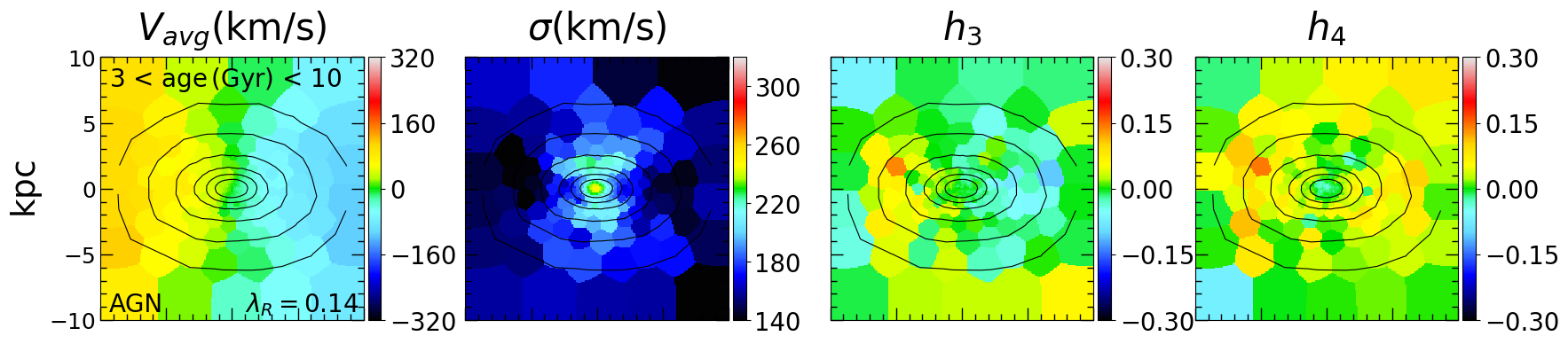}
\includegraphics[width=\textwidth]{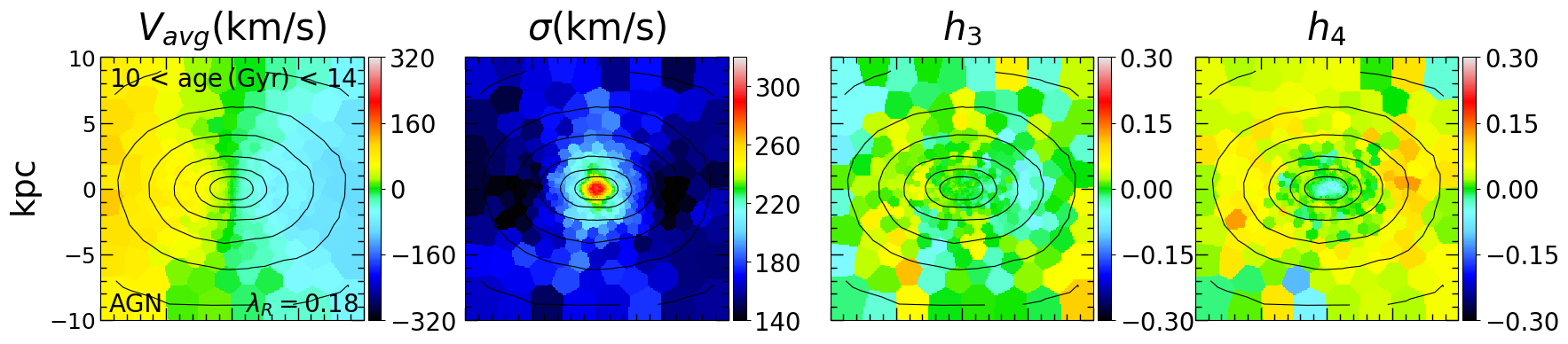}

\caption{Kinematic maps (\vavg,\vdisp,\hthree,\hfour) for age-selected 
stellar components of galaxy 0227 simulated with AGN. From top to bottom, 
stars between 0 and 10 Gyr old and stars older than 10 Gyr.}
\label{fig:kinageagn}
\end{figure*}

Figures \ref{fig:kinagenoagn} and \ref{fig:kinageagn} show the 
kinematic maps of galaxy 0227 at $z=0$ separating the stars between 
different age groups: younger than 3 Gyr (formed after the major 
merger at $z=0.25$), between 3 and 10 Gyr old, and older than 10 Gyr. 
In the case with AGN feedback the galaxy has too few stars belonging 
to the first group, so we skipped it. Without AGN feedback there are 
instead many stars in this group, and they are almost all on rotational 
orbits in a thin disc, with few stars above or below its plane. This disc 
rotates very fast, at around $400 \rm km/s$, has very little velocity 
dispersion and shows quite extreme signatures in the \hthree\ and \hfour\ maps. 
Interestingly the intermediate age group shows an extended velocity dispersion 
signature, perhaps due to the presence of several non-aligned remnants of 
discs that were also quenched in the case with AGN feedback. The older stars 
behave in a relatively similar way with or without AGN feedback. Their 
rotational velocity is smaller and their dispersion much higher, similarly 
to the maps for the accreted component 
(Top panel of Fig. \ref{fig:kininsituagn}). 
From the youngest to the oldest age group, the value of the \lambdar\ 
parameter is 0.84, 0.38 and 0.35 for the case without AGN, and 0.10 and 
0.13 for the case with AGN. The old stars rotate faster in the case 
without AGN, likely because of the deeper central potential.

\bsp

\label{lastpage}

\end{document}